\documentclass[11pt]{article}
\usepackage{fourier}

\usepackage[T1]{fontenc}
\usepackage[latin9]{inputenc}
\usepackage[a4paper]{geometry}
\usepackage[active]{srcltx}
\usepackage{booktabs}
\usepackage{amsmath}
\usepackage{amsthm}
\usepackage{graphicx}
\usepackage[authoryear]{natbib}

\makeatletter

\providecommand{\tabularnewline}{\\}

\@ifundefined{date}{}{\date{}}

\usepackage{amssymb}
\usepackage[small]{titlesec}
\usepackage[font=small,labelfont=bf]{caption}
\usepackage[flushleft]{threeparttable}

\@ifundefined{showcaptionsetup}{}{%
 \PassOptionsToPackage{caption=false}{subfig}}
\usepackage{subfig}
\makeatother

\begin{document}
\title{Free-Riding for Future: Field Experimental Evidence of Strategic Substitutability
in Climate Protest\thanks{The research for this article has been funded by the Deutsche Forschungsgemeinschaft
(DFG, German Research Foundation) under Germany's Excellence Strategy,
cluster EXC 2037 \textquotedblleft CLICCS: Climate, Climatic Change,
and Society\textquotedblright{} (project number: 390683824), and contributes
to the Center for Earth System Research and Sustainability (CEN) of
Universität Hamburg. Henrike Schwickert gratefully acknowledges funding
by the Konrad-Adenauer-Foundation. Data were collected in collaboration
with the Munich office of The Kantar Group Ltd. (London, UK) under
commercial contract. The study was pre-registered in the Randomized
Controlled Trial Registry of the American Economic Association under
code AEARCTR-0004583. All data and materials of the study are freely
available online at the Open Science Framework (OSF) under DOI 10.17605/OSF.IO/Z2EWS.
Thanks to Jane Torbert for preparation of the questionnaire transcript.
The authors declare no conflicts of interest. }}
\author{Johannes Jarke-Neuert\thanks{Center of Earth System Research and Sustainability (CEN), University
of Hamburg. Corresponding author. Mail: Grindelberg 5, 20144 Hamburg,
Germany. Phone: +49 040 42838 8369. E-mail: johannes.jarke-neuert@uni-hamburg.de.}\qquad{}Grischa Perino\thanks{Department of Socioeconomics and Center of Earth System Research and
Sustainability (CEN), University of Hamburg.}\qquad{}Henrike Schwickert\thanks{Department of Socioeconomics, University of Hamburg.}}
\maketitle
\begin{abstract}
We test the hypothesis that protest participation decisions in an
adult population of potential climate protesters are interdependent.
Subjects ($n=1,510$) from the four largest German cities were recruited
two weeks before protest date. We measured participation (ex post)
and beliefs about the other subjects' participation (ex ante) in an
online survey, used a randomized informational intervention to induce
exogenous variance in beliefs, and estimated the causal effect of
a change in belief on the probability of participation using a control
function approach. Participation decisions are found to be strategic
substitutes: a one percentage-point increase of belief causes a .67
percentage-point decrease in the probability of participation in the
average subject. \medskip{}

\textbf{\textit{Keywords:}} collective action; social movement; protest;
environment; climate action; strategic interaction; experiment; causal
mediation; instrumental variable regression\medskip{}

\textbf{\textit{JEL classification:}} C93, D71, D74, D83, Q54
\end{abstract}

\section{Introduction\label{sec:Introduction}}

Political protest is surging at an historically unprecedented level
\citep{Weibel.2015,Almeida.2019,Brannen.2020}, catalyzed by social
media and other digital applications \citep{Jost.2018,Freelon.2020}.
As protest movements play a key role in the process of social change
\citep{Tilly.1978,Acemoglu.2006,Markoff.2014,ChaseDunn.2020}, an
understanding of the former furthers our understanding of the latter.

The intellectual challenge has attracted great interest in various
disciplines (see Section \ref{sec:Discussion} for a brief discussion).
Economic and game theoretic reasoning has contributed the ``strategic
mobilization hypothesis'' \citep{JarkeNeuert.2021}: individual protest
participation decisions are rational and \emph{interdependent}, and
so \emph{beliefs} about others' behavior play a key role in protest
dynamics. However, albeit the hypothesis traces back to at least the
1950s, it has not been credibly tested empirically until very recently---with
mixed results \citep{Cantoni.2019,Manacorda.2020,Gonzalez.2020}---and
never in one of the most prominent movements of our time: climate
protest. This is what we do in the present paper.

Climate change is the realm in which deep social change is most desperately
needed. The scientific facts are on the table \citep{IPCC.2021,Sognnaes.2021},
and a global majority expresses support for change \citep{UNEP.2021}.
Yet, it is not happening at a magnitude that is even close to enough
for prevention of catastrophe \citep{Ripple.2021,CLICCS.2021,Liu.2021}.
It is the youth that is taking worry, anger, and frustration about
the state of matters to the streets \citep{Henry.2020,deMoor.2020},
engaging in ``do-it-ourselves politics'' \citep{Pickard.2019}.\footnote{\label{fn:Young-cohorts-are}Young cohorts are affected by the effects
of global warming mostly, while being under-represented and lacking
agency in the formal political and governmental institutions \citep{Norris.2002,Martin.2012,Sloam.2013,Sloam.2016,Grasso.2016},
such that the incentive to protest is strong \citep{Weiss.2020}.
Furthermore, there are powerful social barriers in the way of effective
climate action \citep{Bernauer.2013,Gifford.2011}, and at least some
are related to the age structure at key loci of decision-making \citep{Aldy.2012,Rickards.2014,Andor.2019}.
Not surprisingly, inter-generational justice is one of the key themes
in climate protest \citep{Holmberg.2019,Hayes.2021,Zabern.2021}} Youth-driven climate protests emerged around UN Climate Summits in
the 2000s and gained significant momentum since the global \textquotedblleft Rise
for Climate\textquotedblright{} campaign and the birth of the ``Fridays
for Future'' movement (FFF) in 2018 \citep{Almeida.2019b,Beckh.2022}.
Other movements (like ``Extinction Rebellion'' or the US-based ``Sunrise
Movement'') and a series of so-called ``Global Climate Strikes''
followed in 2019. The latter's third edition in September was the
largest climate protest event in history, mobilizing reportedly 7.6
million in more than 6,000 events spread across 185 countries \citep{ChaseDunn.2020}.
This event is the context of our study.

We exploited the unique opportunity that adults were explicitly invited
to the Climate Strike events for the first time. The climate protest
movement can apparently mobilize large crowds of youth, and there
is a growing body of evidence on the motivational structure to turn
to the streets (see Section \ref{sec:Discussion}), but much less
is known about the mobilization potential and motivational structure
in the older cohorts of the population. However, this potential is
a key driver of the impact capacity of the movement as a whole, as
the adult population is clearly pivotal for creating sufficient pressure
in the political process (see note \ref{fn:Young-cohorts-are}). 

We recruited more than 1,500 adults in the four largest German cities---Berlin,
Hamburg, Munich, and Cologne---two weeks before publicly announced
local protest events that happened simultaneously on September 20,
2019. We measured participation (ex post) and beliefs about the other
subjects' participation (ex ante) with an online survey, used a randomized
informational intervention to induce exogenous variance in beliefs,
and estimated the causal effect of a change in belief on the probability
of participation using a control function approach. The data clearly
support the strategic mobilization hypothesis, and we find participation
decisions to be strategic substitutes.

The remainder of the paper is structured as follows. In Section \ref{sec:Hypothesis}
we formulate the testable hypothesis. The data and the collection
procedures are desribed in Section \ref{sec:Data}. Section \ref{sec:Methods}
explains the statistical inference methods. The empirical results
and evidence supporting the identifying assumptions are presented
in Section \ref{sec:Results}. Section \ref{sec:Discussion} discusses
the contribution to the literature, and in Section \ref{sec:Conclusion}
we conclude by daring an outlook regarding the future of climate protest,
and by suggesting avenues for further research.

\section{Hypothesis\label{sec:Hypothesis}}

We draw on a micro-founded parametric model of political protest mobilization
\citep{JarkeNeuert.2021}, which explains the mean participation probability
in a population of potential protesters as a function of preferences
and beliefs. Specifically, the conditional probability of protest
participation is assumed to be governed by
\begin{equation}
\mathrm{Pr}\left(a=1\mid b,\boldsymbol{x}\right)=\Phi\left(\alpha+\beta\cdot b+\boldsymbol{x}\cdot\boldsymbol{\gamma}\right)\label{eq:ProbitModel}
\end{equation}
where $a\in\left\{ 0,1\right\} $ indicates protest participation,
$b\in\left[0,1\right]$ is the probabilistic belief regarding the
participation probability of an average other potential participant,
$\boldsymbol{x}=\left(x_{1},\ldots,x_{k}\right)$ is a vector of optional
covariates, $\Phi$ is the standard normal cumulative distribution
function, and $\left(\alpha,\beta,\gamma_{1},\ldots,\gamma_{k}\right)\in\mathbb{R}^{k+2}$
are fixed parameters. Parameter $\beta$ captures the interdependent
(``strategic'') component of the participation decision, with values
$\beta>0$ indicating \emph{strategic complementarity} (a potential
participant's marginal utility of participation is increasing in expected
protest size) and $\beta<0$ indicating \emph{strategic substitutability}
(marginal utility is decreasing in expected protest size).\footnote{The connection to the conventional definitions of strategic complementarity
and substitutability, respectively, in terms of marginal utility is
made explicit in \citet{JarkeNeuert.2021} via the latent variable
formulation of \eqref{eq:ProbitModel}, which is derived from a simple
random utility model of decision-making. Mild and plausible normality
assumptions about the distribution of random utilities justify the
probit link function. It also follows that parameter $\alpha$ and
the latent variable model error captures motivations to participate
(or not) that are independent from turnout (e. g. moral duty, see
Section \ref{sec:Discussion} for evidence).} The basic aim of the present study is a test of hypothesis
\begin{quote}
$\mathrm{H}_{0}$: $\beta=0$ vs. $\mathrm{H}_{1}$: $\beta\neq0$.
\end{quote}
We also want to estimate the sign and the magnitude of the marginal
effect 
\[
\frac{\partial\mathrm{Pr}\left(a=1\mid b,\boldsymbol{x}\right)}{\partial b}=\varphi\left(\alpha+\beta\cdot b+\boldsymbol{x}\cdot\boldsymbol{\gamma}\right)\cdot\beta
\]
evaluated at a suitable value of belief and averaged over $\boldsymbol{x}$
($\varphi$ denoting the standard normal density), the average partial
effect (APE) of a change of belief on the probability of participation.

In principle, the parameters $\left(\alpha,\beta,\boldsymbol{\gamma}\right)$
and their asymptotic variance can be estimated from any dataset containing
paired (independent and identically distributed) observations of participation
decisions and beliefs by (conditional) maximum likelihood estimation
\citep[pp. 389]{Wooldridge.2010}. However, beliefs are not only difficult
to measure (as they are cognitive constructs), but they are also influenced
by a myriad of uncontrollable and unobservable (hence not included
in $\boldsymbol{x}$) events, such that estimates of the parameters
based on purely observational data would likely be polluted by spurious
correlation or masked causation.\footnote{Technically, if beliefs are not independent from the latent variable
model error (i. e. $b$ is not exogenous), which is likely the case
with purely observational data, then the conditional densities on
which the likelihood maximization problem is defined will be misspecified,
and estimates would hence be inconsistent. See \citet[pp. 391]{Wooldridge.2010}
for details.} Consequently, we could falsely reject or falsely not reject $\mathrm{H}_{0}$.
We therefore resort to a carefully designed randomized controlled
trial design involving an information provision intervention (section
\ref{sec:Data}), and advanced methods of data analysis (section \ref{sec:Methods}).

\section{Data\label{sec:Data}}

The study was built around the so-called Third Global Climate Strike
on September 20, 2019, a collection of protest events at many locations
around the globe that were coordinated with respect to date but organized
locally by climate activist volunteers (typically FFF activists). 

We focused on the events in four major German cities---Berlin, Hamburg,
Munich, and Cologne---to exploit two critical features.\footnote{The four cities are the largest (in terms of population) in Germany.
Population sizes, gender distributions, and age-class distributions
as of end 2019 from official census records are provided in Table
\ref{tab:PopProperties} in the Appendix.} First, they were not spontaneous, but public calls with a specific
date, time, and location were circulated by the local FFF organizers
weeks ahead of time via various means (flyers, stickers, social media,
conventional media, etc.).\footnote{The events were scheduled on 12pm at Brandenburger Tor in Berlin,
12pm at Jungfernstieg in Hamburg, 12pm at Königsplatz in Munich, and
11am at Hans-Böckler-Platz in Cologne.} This was essential for planning a study around the events. Second,
adults at all ages were explicitly invited by the organizers to participate.
This feature was new to the Climate Strike movement that has previously
been driven by youth \citep{deMoor.2020}, and it provided not only
a unique opportunity to study ``a game in the making'' within a
new segment of the population, but the limited experience of those
new ``players'' with the ``game'' granted us scope for an informational
intervention, to be described in section \ref{subsec:Intervention}.
Before that we expose the sampling procedures in Section \ref{subsec:Sampling},
and show in Section \ref{subsec:Measurement} how the variables of
the participation model have been measured. 

\subsection{Sampling\label{subsec:Sampling}}

Data were collected in collaboration with the Munich office of The
Kantar Group Ltd. (London, UK) under commercial contract. The company
administers local opt-in online panels of volunteers (typically used
for market research) in each of the four subject cities. 

Invitations to participate in a scientific study on ``environmental
and climate protection matters'' involving three sequential surveys
were sent out to registered panelists aged between 18 and 69 by e-mail
on September 6, 2019. They were informed that they would be compensated
financially for each completed survey according to Kantar's default
lump-sum rates,\footnote{We deliberately opted against response-conditioned incentives to avoid
undesirable side-effects. The data collection contractor (Kantar)
informed us that they have established a ``code of honor'' for truthfulness
with their respondents, and response-conditioned incentives would
undermine this code and make non-truthfulness salient in the first
place. This is supported by evidence \citep{Gritz.2004,Kamenica.2012}.
There is also evidence that scoring rules for belief elicitation can
adversely affect accuracy and induce hedging, although the problem
appears to hinge on specific details of implementation \citep{Gaechter.2003,Blanco.2010,Armantier.2013,Schlag.2015}.} plus a bonus for completing all three surveys. The full invitation
text is provided in the OSF online materials \citep{JarkeNeuert.2021mat}. 

Panelists could accept the invitation and complete the first survey
between September 6 (Friday) and September 11 (Wednesday) at midnight,
local time. A total of 2,576 subjects accepted the invitation, two
dropped out during the first survey. Those 2,574 subjects who completed
the first survey were invited to the second survey, which was open
between September 16 (Monday) and September 20 (Friday) at noon, local
time. The 1,879 panelists that completed the second survey were invited
to the third survey that was fielded between December 5 (Thursday)
and December 16 (Monday).\footnote{The third survey has been originally planned for the period between
September 21 and October 1 but had to be postponed due to technical
problems of the data collection contractor (Kantar). Since the single
survey instrument used for the experiment was a simple fact question
about participation in the local protest event (see section \ref{subsec:Measurement}),
it is unlikely that the delay caused any kind of problem. We used
the opportunity to also elicit participation decisions for the so-called
Fourth Global Climate Strike (November 28, 2019) for explorative purposes.} A total of 1,510 subjects completed all three surveys. A breakdown
of the sampling process by location is shown in Table \ref{tab:FullSmp}.
The sample in the bottom row, comprising all subjects that completed
all three surveys, enters the data analysis. Sample breakdowns by
location, gender and age are provided in the Appendix in Table \ref{tab:SampleProperties}.\footnote{Probability-expected sampling frequencies based on the gender and
age distributions in the local populations of inhabitants are listed
in Table \ref{tab:Diagnostics} in the Appendix. They show that the
oldest age group (65 or older) is under-represented somewhat in our
sample, which is to be expected by the sampling restriction (69 or
younger) and from online access panels in general \citep{Blasius.2010}.
To compensate, the next younger age group (50-65) is over-sampled
a bit. Overall, the sample approximates the local populations acceptably,
although we emphasize that representativeness was not an objective
for this study.}
\begin{table}[t]
\caption{Sample breakdown by location and survey.\label{tab:FullSmp}}

\centering{}%
\begin{tabular*}{1\textwidth}{@{\extracolsep{\fill}}lrrrrr}
\toprule 
 & \multicolumn{1}{c}{Berlin} & \multicolumn{1}{c}{Hamburg} & \multicolumn{1}{c}{Munich} & \multicolumn{1}{c}{Cologne} & \multicolumn{1}{c}{Overall}\tabularnewline
\midrule
1st survey & $848$ $\left(.3294\right)$ & $651$ $\left(.2529\right)$ & $490$ $\left(.1904\right)$ & $585$ $\left(.2273\right)$ & $2,574$\tabularnewline
2nd survey & $610$ $\left(.3246\right)$ & $481$ $\left(.2560\right)$ & $363$ $\left(.1932\right)$ & $425$ $\left(.2262\right)$ & $1,879$\tabularnewline
3rd survey & $490$ $\left(.3245\right)$ & $399$ $\left(.2642\right)$ & $280$ $\left(.1854\right)$ & $341$ $\left(.2258\right)$ & $1,510$\tabularnewline
\bottomrule
\end{tabular*}\textit{\footnotesize{}\medskip{}
}\textit{\scriptsize{}Table notes:}{\scriptsize{} Listed are the counts
of subjects that completed the respective survey in the respective
location class, with row percentages in parentheses. Kruskal-Wallis
equality-of-populations rank tests do not reject the hypothesis that
sample attrition between the first and the second survey ($\chi^{2}\left(3\right)=1.075$
with ties, $p=.7831$) and between the second and the third ($\chi^{2}\left(3\right)=2.623$
with ties, $p=.4534$) is equal across locations. Overall fractional
attrition rates are $.270$ for the second survey (relative to the
first) and $.196$ for the third survey (relative to the second).}{\scriptsize\par}
\end{table}

\subsection{Measurement instruments\label{subsec:Measurement}}

\begin{figure}[b]
\caption{Schematic overview of the data generation process.\label{fig:Schema}}

\centering{}\includegraphics[width=0.7\textwidth]{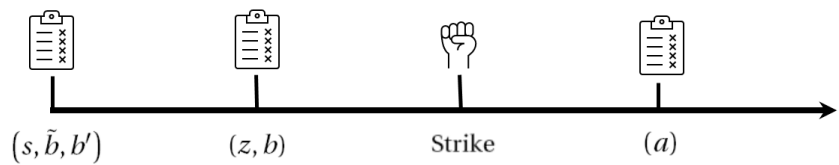}
\end{figure}

In the three surveys we measured the variables of the model described
equation \eqref{eq:ProbitModel} plus additional data. A schematic
overview is shown in Figure \ref{fig:Schema}. At the beginning of
the first questionnaire, respondents were briefly introduced to the
subject matter in objective and neutral language, and confronted with
the public call applicable to their city of residence. It was also
explained to them that people between 18 and 69 years living in the
same city are surveyed, and that this group approximates the structure
of the local population. We describe measurement of the key variables
here, the full questionnaires in German (original) and English (transcript)
are available in the OSF online materials \citep{JarkeNeuert.2021mat}.\footnote{There are many additional socio-demographic and opinion items and
a finer treatment (separating a ``younger'' and an ``older'' age
class) in the questionnaire (and the dataset), which we donate to
the community for further research. We devoted great care in asking
the questions in a sequence that does not pollute our key instruments.} 

The participation indicator $a\in\left\{ 0,1\right\} $ was measured
in the third survey with the following five-point nominal-scale instrument
(Q25 in the questionnaire):\footnote{We used the five-point distinction instead of asking a binary question
to avoid misunderstandings of what ``participating'' means. We wanted
be able to distinguish between actual participants (response 1) and
people that happened to be there for other reasons (responses 2 and
3). Since there were events at many locations on the same day, we
also wanted to distinguish participants in the local event (response
1) and other events (response 4). The questionnaire contained a number
of additional questions for explorative purposes. The questionnaire
also referred to ``Fridays for Future rally'' because the movement
is commonly known under this term in Germany, and the call was published
under that label.}
\begin{verse}
\texttt{\small{}In the first survey, you were asked about your intention
to attend the \textquotedblleft Fridays for Future\textquotedblright{}
rally on September 20, 2019, in {[}applicable city{]}. Did you participate
in the rally in {[}applicable city{]} and if so, in what way?}{\small\par}

\texttt{\small{}1. Yes, I was there as a participant.}{\small\par}

\texttt{\small{}2. Yes, I was there as an observer. }{\small\par}

\texttt{\small{}3. Yes, I was there as a counter-demonstrator. }{\small\par}

\texttt{\small{}4. No, I was not there, but at a different ``Fridays
for Future'' event that day. }{\small\par}

\texttt{\small{}5. No, I did not participate in any ``Fridays for
Future'' event that day.}{\small\par}
\end{verse}
The participation indicator was set to $a=1$ for subjects that selected
response 1, and to $a=0$ otherwise. The distributions of responses
at each location and in the pooled sample is shown in Table \ref{tab:SumPart}.
\begin{table}[t]
\caption{Measured behavior with regard to the climate protest on September
20, 2019.\label{tab:SumPart}}

\centering{}%
\begin{tabular*}{1\textwidth}{@{\extracolsep{\fill}}lrrrrr}
\toprule 
 & \multicolumn{1}{c}{Berlin} & \multicolumn{1}{c}{Hamburg} & \multicolumn{1}{c}{Munich} & \multicolumn{1}{c}{Cologne} & \multicolumn{1}{c}{Overall}\tabularnewline
\midrule
Q25.1 Local participant ($a=1)$ & $.0918$ & $.1128$ & $.1036$ & $.1378$ & $.1099$\tabularnewline
Q25.2 Local observer & $.0735$ & $.1153$ & $.1000$ & $.0880$ & $.0927$\tabularnewline
Q25.3 Local counter-protester & $.0000$ & $.0000$ & $.0000$ & $.0029$ & $.0007$\tabularnewline
Q25.4 Participant elsewhere & $.0245$ & $.0401$ & $.0321$ & $.0411$ & $.0338$\tabularnewline
Q25.5 Absent & $.8102$ & $.7318$ & $.7643$ & $.7302$ & $.7629$\tabularnewline
\bottomrule
\end{tabular*}\textit{\footnotesize{}\medskip{}
}\textit{\scriptsize{}Table notes:}{\scriptsize{} Listed are the empirical
distributions of responses to survey instrument Q25 in the third survey,
which is equal to the distributions in the final sample (subjects
that completed all three surveys, the bottom row of Table \ref{tab:FullSmp}),
in the form of relative frequencies. A Kruskal-Wallis equality-of-populations
rank test rejects equality of distributions across locations at a
five percent level of significance ($\chi^{2}\left(3\right)=9.647$
with ties, $p=.0218$).}{\scriptsize\par}
\end{table}

The belief $b$ was elicited (post-intervention) in the second survey,
a pre-intervention belief that helps in controlling for the endogeneity
of $b$ (see Section \ref{sec:Methods})---to be denoted $b^{\prime}$---was
measured in the first survey. The latter was elicited with the following
input-box instrument (Q13 in the questionnaire): 
\begin{verse}
\texttt{\small{}What do you think \textemdash{} what percentage of
all survey respondents will actually participate in the \textquotedblleft Fridays
for Future\textquotedblright{} rally on September 20, 2019, in {[}applicable
city{]}? There is no right or wrong answer, we are interested in your
personal assessment. I estimate that of all survey respondents, \_\_\_.\_
percent will actually participate.}{\small\par}
\end{verse}
The post-intervention belief instrument (Q19) in the second survey
was essentially identical, the only difference was an introductory
sentence that reminded subjects of their own previous response in
Q13, and then continued to explain:
\begin{verse}
\texttt{\small{}Perhaps your estimate of the proportion of respondents
who will actually participate in the rally has changed since the last
survey. Please estimate this value again now.}\texttt{ }
\end{verse}
Respondents could type in a number between 0 and 100 with a resolution
of a single decimal digit, respectively, which was re-scaled to the
unit interval for analysis as variables $\left(b^{\prime},b\right)\in\left[0,1\right]^{2}$.
Summary statistics of the empirical distributions at each location
and in the pooled sample are listed in Tables \ref{tab:SumPriors}
and \ref{tab:SumPosteriors}, respectively.
\begin{table}
\caption{Summary of pre-intervention beliefs regarding others' participation
($b^{\prime}$) in the climate protest on September 20, 2019.\label{tab:SumPriors}}

\centering{}%
\begin{tabular*}{1\textwidth}{@{\extracolsep{\fill}}lrrrrr}
\toprule 
 & \multicolumn{1}{c}{Berlin} & \multicolumn{1}{c}{Hamburg} & \multicolumn{1}{c}{Munich} & \multicolumn{1}{c}{Cologne} & \multicolumn{1}{c}{Overall}\tabularnewline
\midrule
5-percentile & $.020$ $\left(.020\right)$ & $.020$ $\left(.020\right)$ & $.020$ $\left(.020\right)$ & $.020$ $\left(.020\right)$ & $.020$ $\left(.020\right)$\tabularnewline
25-percentile & $.100$ $\left(.100\right)$ & $.100$ $\left(.100\right)$ & $.080$ $\left(.080\right)$ & $.090$ $\left(.100\right)$ & $.100$ $\left(.100\right)$\tabularnewline
Median & $.200$ $\left(.200\right)$ & $.200$ $\left(.200\right)$ & $.200$ $\left(.200\right)$ & $.200$ $\left(.200\right)$ & $.200$ $\left(.200\right)$\tabularnewline
75-percentile & $.400$ $\left(.400\right)$ & $.350$ $\left(.400\right)$ & $.350$ $\left(.360\right)$ & $.400$ $\left(.400\right)$ & $.370$ $\left(.400\right)$\tabularnewline
95-percentile & $.700$ $\left(.700\right)$ & $.670$ $\left(.660\right)$ & $.600$ $\left(.650\right)$ & $.650$ $\left(.700\right)$ & $.660$ $\left(.700\right)$\tabularnewline
\midrule
Mean & $.2600$ $\left(.2698\right)$ & $.2473$ $\left(.2570\right)$ & $.2381$ $\left(.2429\right)$ & $.2481$ $\left(.2615\right)$ & $.2499$ $\left(.2596\right)$\tabularnewline
Standard deviation & $.2195$ $\left(.2260\right)$ & $.1954$ $\left(.2025\right)$ & $.1917$ $\left(.2012\right)$ & $.2080$ $\left(.2217\right)$ & $.2056$ $\left(.2148\right)$\tabularnewline
\bottomrule
\end{tabular*}\textit{\footnotesize{}\medskip{}
}\textit{\scriptsize{}Table notes:}{\scriptsize{} Listed are statistics
of the empirical distributions in the final sample (subjects that
completed all three surveys, the bottom row of Table \ref{tab:FullSmp}),
with statistics of the distributions in the intermediate first survey
sample (including subjects that completed only the first or the first
two surveys, the top row of Table \ref{tab:FullSmp}) in parentheses.
Two-sample Kolmogorov-Smirnov tests do not reject equality of distribution
functions in the final sample and the class of dropouts (Berlin $D=.0774$,
exact $p=.157$; Hamburg $D=.0894$, exact $p=.158$; Munich $D=.0429$,
exact $p=.974$; Cologne $D=.0661$, exact $p=.535$). A Kruskal-Wallis
equality-of-populations rank test does not reject equality of distributions
across locations ($\chi^{2}\left(3\right)=1.109$ with ties, $p=.7748$).
A Shapiro-Wilk test rejects normality ($z=11.181$, $p=.0000$). A
maximum likelihood fit to a beta distribition yields a shape parameter
estimate of $.9805$ (SEE $.0317$, $p=.000)$ and a scale parameter
estimate of $2.8988$ (SEE $.1093$, $p=.000)$ at log $\mathcal{L}=631.82672$.
A one-sample Kolmogorov-Smirnov test does not reject equality between
this theoretical distribution and the empirical distribution ($D=.0917$,
$p=.000$). }{\scriptsize\par}
\end{table}
\begin{table}
\caption{Summary of post-intervention beliefs regarding others' participation
($b$) in the climate protest on September 20, 2019.\label{tab:SumPosteriors}}

\centering{}%
\begin{tabular*}{1\textwidth}{@{\extracolsep{\fill}}lrrrrr}
\toprule 
 & \multicolumn{1}{c}{Berlin} & \multicolumn{1}{c}{Hamburg} & \multicolumn{1}{c}{Munich} & \multicolumn{1}{c}{Cologne} & \multicolumn{1}{c}{Overall}\tabularnewline
\midrule
5-percentile & $.020$ $\left(.030\right)$ & $.030$ $\left(.030\right)$ & $.030$ $\left(.030\right)$ & $.030$ $\left(.020\right)$ & $.030$ $\left(.020\right)$\tabularnewline
25-percentile & $.100$ $\left(.110\right)$ & $.100$ $\left(.100\right)$ & $.105$ $\left(.100\right)$ & $.100$ $\left(.100\right)$ & $.100$ $\left(.100\right)$\tabularnewline
Median & $.235$ $\left(.250\right)$ & $.200$ $\left(.210\right)$ & $.250$ $\left(.220\right)$ & $.230$ $\left(.230\right)$ & $.230$ $\left(.200\right)$\tabularnewline
75-percentile & $.350$ $\left(.400\right)$ & $.350$ $\left(.367\right)$ & $.400$ $\left(.400\right)$ & $.350$ $\left(.350\right)$ & $.350$ $\left(.400\right)$\tabularnewline
95-percentile & $.650$ $\left(.660\right)$ & $.600$ $\left(.600\right)$ & $.600$ $\left(.600\right)$ & $.600$ $\left(.600\right)$ & $.600$ $\left(.700\right)$\tabularnewline
\midrule
Mean & $.2647$ $\left(.2724\right)$ & $.2568$ $\left(.2607\right)$ & $.2587$ $\left(.2581\right)$ & $.2556$ $\left(.2588\right)$ & $.2594$ $\left(.2636\right)$\tabularnewline
Standard deviation & $.1982$ $\left(.2003\right)$ & $.1784$ $\left(.1831\right)$ & $.1768$ $\left(.1801\right)$ & $.1845$ $\left(.1855\right)$ & $.1860$ $\left(.1888\right)$\tabularnewline
\bottomrule
\end{tabular*}\textit{\footnotesize{}\medskip{}
}\textit{\scriptsize{}Table notes:}{\scriptsize{} Listed are statistics
of the empirical distributions in the final sample (subjects that
completed all three surveys, the bottom row of Table \ref{tab:FullSmp}),
with statistics of the distributions in the intermediate second survey
sample (including subjects that completed the first two surveys, the
middle row of Table \ref{tab:FullSmp}) in parentheses. Two-sample
Kolmogorov-Smirnov tests do not reject equality of distribution functions
in the final sample and the class of dropouts in Hamburg ($D=.1109$,
exact $p=.343$), Munich ($D=.0708$, exact $p=.874$) and Cologne
($D=.0625$, exact $p=.937$), but in Berlin at a five percent level
of significance ($D=.1464$, exact $p=.028$). A Kruskal-Wallis equality-of-populations
rank test does not reject equality of distributions across locations
($\chi^{2}\left(3\right)=.215$ with ties, $p=.9751$). A Shapiro-Wilk
test rejects normality ($z=10.562$, $p=.0000$). A maximum likelihood
fit to a beta distribition yields a shape parameter estimate of $1.2772$
(SEE $.0422$, $p=.000)$ and a scale parameter estimate of $3.6295$
(SEE $.1353$, $p=.000)$ at log $\mathcal{L}=631.82672$, with a
one-sample Kolmogorov-Smirnov test again not rejecting equality between
the theoretical and empirical distributions ($D=.0538$, $p=.000$).}{\scriptsize\par}
\end{table}

\subsection{Intervention design\label{subsec:Intervention}}

The intervention design is adapted from \citet{Cantoni.2019}. Two
survey instruments relating to \emph{intentions} to participate in
the protest were included in the first questionnaire \emph{before}
the pre-intervention belief $b^{\prime}$ was elicited. They served
as inputs for the information treatment given in the second survey.
The first was the following simple four-point nominal-scale instrument
(Q9 in the questionnaire):
\begin{verse}
\texttt{\small{}Are you planning to participate in the \textquotedblleft Fridays
for Future\textquotedblright{} rally on September 20, 2019, in {[}applicable
city{]}?}{\small\par}

\texttt{\small{}1. Yes, I do plan to participate. }{\small\par}

\texttt{\small{}2. I am not sure yet, but I rather plan to participate. }{\small\par}

\texttt{\small{}3. I am not sure yet, but I rather not plan to participate. }{\small\par}

\texttt{\small{}4. No, I do not plan to participate.}{\small\par}
\end{verse}
The distribution of responses is shown in the top four rows of Table
\ref{tab:SumIntentions}. The fraction of subjects that selected response
1 (intending to participate) or 2 (rather intending to participate),
was calculated for each city after the first survey was completed.
This variable, denoted $s\in\left[0,1\right]$, was accordingly a
location-specific constant. The values are shown in the fifth row
of Table \ref{tab:SumIntentions}.

The second instrument elicited location-specific beliefs regarding
$s$ (Q10 in the questionnaire): 
\begin{verse}
\texttt{\small{}Each survey participant answers the previous question. What
do you think \textemdash{} what percentage of all respondents answers
the previous question number 9 with \textquotedblleft Yes, I do plan
to participate\textquotedblright{} or \textquotedblleft I am not sure
yet, but I rather plan to participate\textquotedblright . There is
no right or wrong answer, we are interested in your personal assessment. I
estimate that of all survey respondents, \_\_\_.\_ percent answers
the previous question number 9 with \textquotedblleft Yes, I do plan
to participate\textquotedblright{} or \textquotedblleft I am not sure
yet, but I rather plan to participate''.}{\small\par}
\end{verse}
Again, respondents could type in a number between 0 and 100 at a resolution
of a single decimal digit, that was re-scaled to the unit interval
for analysis as variable $\tilde{b}\in\left[0,1\right]$. Means and
standard deviations of the empirical distributions are listed in the
bottom row of Table \ref{tab:SumIntentions}.
\begin{table}[t]
\caption{Summary of intervention input data: signal values ($s$) and reference
beliefs ($\tilde{b}$).\label{tab:SumIntentions}}

\centering{}%
\begin{tabular*}{1\textwidth}{@{\extracolsep{\fill}}lrrrrr}
\toprule 
 & \multicolumn{1}{r}{Berlin} & \multicolumn{1}{r}{Hamburg} & \multicolumn{1}{r}{Munich} & \multicolumn{1}{r}{Cologne} & \multicolumn{1}{r}{Overall}\tabularnewline
\midrule
Q9.1 Yes & $.0660$ & $.0707$ & $.0592$ & $.0838$ & $.0699$\tabularnewline
Q9.2 Rather yes & $.2594$ & $.2965$ & $.3082$ & $.2821$ & $.2832$\tabularnewline
Q9.3 Rather no & $.1946$ & $.2166$ & $.2102$ & $.1641$ & $.1962$\tabularnewline
Q9.4 No & $.4800$ & $.4163$ & $.4224$ & $.4701$ & $.4507$\tabularnewline
\midrule
$s$ & $.325$ & $.367$ & $.367$ & $.366$ & \tabularnewline
\midrule
Mean of $\tilde{b}$ & $.3379\pm.2339$ & $.3315\pm.2246$ & $.3240\pm.2295$ & $.3324\pm.2381$ & $.3324\pm.2317$\tabularnewline
\bottomrule
\end{tabular*}\textit{\footnotesize{}\medskip{}
}\textit{\scriptsize{}Table notes:}{\scriptsize{} The top four rows
list the empirical distributions of responses to the participation
intention instrument (Q9) in the first survey sample (subjects that
completed the first survey, the top row of Table \ref{tab:FullSmp})
in the form of relative frequencies. A Kruskal-Wallis equality-of-populations
rank test does not reject equality of distributions across locations
($\chi^{2}\left(3\right)=5.926$ with ties, $p=.1153$). The fifth
row shows the location-specific signal values ($s$) calculated from
adding the relative frequencies of the first two rows (Q9.1 and Q9.2),
respectively. Signal values were shown to treated subjects as a percentage
with a single decimal digit. The bottom row lists the means $\pm$
standard deviations of beliefs regarding participation intentions
($\tilde{b}$), elicited by instrument Q10. A Kruskal-Wallis equality-of-populations
rank test does not reject equality of distributions across locations
($\chi^{2}\left(3\right)=1.250$ with ties, $p=.7411$).}{\scriptsize\par}
\end{table}

At the beginning of the second survey, subjects were randomly assigned
to either a treatment condition, indicated by $z=1$, or a control
condition ($z=0$). A breakdown of the sample by experimental condition
is shown in Table \ref{tab:Assignment}. Notably, a two-sided two-sample
test of proportions does not reject equality of assignment proportions
in the final sample and the class of subjects that dropped out after
the second survey ($.6795$ vs. $.6748$, $z=-.1723$, $p=.8632$,
see the Table \ref{tab:Assignment} notes for location-specific tests),
which supports the assumption that attrition is independent from treatment
assignment.
\begin{table}[t]
\caption{Sample breakdown by city and experimental condition.\label{tab:Assignment}}

\centering{}%
\begin{tabular*}{1\textwidth}{@{\extracolsep{\fill}}lrrrrr}
\toprule 
Second survey & \multicolumn{1}{c}{Berlin} & \multicolumn{1}{c}{Hamburg} & \multicolumn{1}{c}{Munich} & \multicolumn{1}{c}{Cologne} & \multicolumn{1}{c}{Overall}\tabularnewline
\midrule
Control ($z=0$) & $190$ $\left(.3115\right)$ & $162$ $\left(.3368\right)$ & $128$ $\left(.3526\right)$ & $124$ $\left(.2918\right)$ & $604$ $\left(.3214\right)$\tabularnewline
Treatment ($z=1$) & $420$ $\left(.6885\right)$ & $319$ $\left(.6632\right)$ & $235$ $\left(.6474\right)$ & $301$ $\left(.7082\right)$ & $1,275$ $\left(.6786\right)$\tabularnewline
\midrule
Final sample &  &  &  &  & \tabularnewline
\midrule
Control ($z=0$) & $147$ $\left(.3000\right)$ & $134$ $\left(.3356\right)$ & $103$ $\left(.3679\right)$ & $100$ $\left(.2933\right)$ & $484$ $\left(.3205\right)$\tabularnewline
Treatment ($z=1$) & $343$ $\left(.7000\right)$ & $265$ $\left(.6642\right)$ & $177$ $\left(.6321\right)$ & $241$ $\left(.7067\right)$ & $1,026$ $\left(.6795\right)$\tabularnewline
\bottomrule
\end{tabular*}\textit{\footnotesize{}\medskip{}
}\textit{\scriptsize{}Table notes:}{\scriptsize{} Listed are the counts
of subjects that completed the first two surveys (top panel) or all
three surveys (bottom panel) in the respective location class, broken
down by experimental condition with column percentages in parentheses.
The target treatment assignment probability was two over three. Two-sided
binomial probability tests do not reject the null hypothesis that
observed assignment (in the final sample) is on target (Berlin: exp.
$k=326.67$, $p=.1251$; Hamburg: exp. $k=266.0$, $p=.9155$; Munich:
exp. $k=186.67$, $p=.2285$; Cologne: exp. $k=227.33$, $p=.1211$;
Overall: exp. $k=1006.67$, $p=.2997$). Two-sided two-sample tests
of proportions do not reject equality of assignment proportions in
the final sample and the class of dropouts (Berlin: $z=-1.2367$,
$p=.2162$; Hamburg: $z=-.0981$, $p=.9218$; Munich: $z=1.1162$,
$p=.2643$; Cologne: $z=.1362$).}{\scriptsize\par}
\end{table}

Subjects in the treatment group were informed about their location-specific
$s$ before the post-intervention belief $b$ was elicited: 
\begin{quotation}
\texttt{\small{}The first survey showed that {[}$100\cdot s${]} percent
of all respondents plan or rather plan to participate.}{\small\par}
\end{quotation}
Subjects in the control group did not receive this information (the
above sentence was just not displayed). Otherwise the experimental
conditions were identical. The idea is that treatment in combination
with the reference belief $\tilde{b}$ induces an informational stimulus
that gives reason to adjust the post-intervention belief $b$ relative
to control. The induced causal effect in beliefs can be exploited
to address the problems described at the end of Section \ref{sec:Hypothesis}
and adequately test $\mathrm{H}_{0}$. Details follow in the next
section.

\section{Statistical methods\label{sec:Methods}}

The basic statistical approach is to use treatment assignment ($z$)
as an instrumental variable for the potentially endogenous belief
variable $b$ \citep{Imbens.1994,Angrist.1996,Clarke.2012}. By randomization,
treatment status is credibly exogenous. Yet, under mild assumptions
it affects beliefs, as we demonstrate in section \ref{subsec:Auxiliary}.
We will then continue to specify a generic parametric model that identifies
average effect of treatment (ATE) on beliefs. In subsection \ref{subsec:Augmented}
we will connect this model to our probit participation model \eqref{eq:ProbitModel}
and show how the treatment-induced variation in post-intervention
beliefs can be exploited to recover the true causal APE and reliably
test $\mathrm{H}_{0}$. The direct test of the hypothesis and an indirect
test of the key identifying assumptions are specified in Section \ref{subsec:Test-of-hypothesis}.

\subsection{Instrumental belief updating model\label{subsec:Auxiliary}}

Drawing on the potential outcomes framework \citep{Rubin.1974,Rubin.1990,Holland.1986},
let $b_{z}$ be the potential belief realized under treatment status
$z$, such that the (individual) causal effect of treatment is by
definition $\tau:=b_{1}-b_{0}$. Given the intervention design described
above, this effect will be some function of the stimulus $\left(s,\tilde{b}\right)$
presented to subjects by treatment. A mild assumption is 
\begin{equation}
\tau=\psi\left(s-\tilde{b}\right)\label{eq:BeliefTheory}
\end{equation}
with $\psi$ being a monotonically increasing function that goes through
the origin.\footnote{The monotonicity assumption is actually essential for the instrumental
variable approach used in this study \citep[see][and the discussion at the end of the section]{Imbens.1994,Angrist.1996}.} It reflects the plausible theory that a treated subject with condition
$\tilde{b}<s$ learns that actually more people are intending to join
the protest than expected, which provides a logical reason for $b_{1}>b_{0}$.
The converse logic holds for $\tilde{b}>s$, whereas under condition
$\tilde{b}=s$ expectations regarding participation intentions turned
out to be accurate, such that there is no reason for $b_{1}\neq b_{0}$.

Depending on treatment assignment, exactly one of the potential post-intervention
beliefs is observed for each subject,
\begin{equation}
b=z\cdot b_{1}+\left(1-z\right)\cdot b_{0}=z\cdot\tau+b_{0}\label{eq:ObsBelief}
\end{equation}
It will be convenient to center the data by defining the \emph{change}
of belief over time relative to the pre-intervention belief, $\Delta b:=b-b^{\prime}$,
such that we observe
\[
\Delta b=z\cdot\tau+\left(b_{0}-b^{\prime}\right)
\]
The latter term in (redundant) parentheses captures any belief adjustment
over time that is not controlled by treatment, such as updates reflecting
uncontrolled information inflow from other sources. It also has the
feature that it is centered about zero, such that $\Delta b>0$ indicates
upwards adjustment (expecting others' participation more likely than
before), and $\Delta b<0$ indicates downwards adjustment (expecting
others' participation less likely than before). This facilitates the
joint analysis of all four location strata and already addresses the
endogeneity issue to some extent, as pre-intervention beliefs serve
as an individual offset.

To account for directional treatment effect heterogeneity implied
by \eqref{eq:BeliefTheory}, define the condition indicator 
\[
c=\begin{cases}
1 & \tilde{b}\geq s\\
0 & \tilde{b}<s
\end{cases}
\]
such that individuals with a reference belief (at or) \emph{above}
the signal value form the ``above group'' ($c=1$), and individuals
with a reference belief \emph{below} signal value form the ``below
group'' ($c=0$). We can then parametrize the average change of belief
conditional on class $c$ and treatment status $z$ by the generalized
linear model (GLM)
\begin{equation}
\mathrm{E}\left(\Delta b\mid c,z\right)=L\left(\theta_{0}+\theta_{1}\cdot z+\theta_{2}\cdot c+\theta_{3}\cdot z\cdot c\right)\label{eq:BeliefGLM}
\end{equation}
where $L$ denotes the link function and $\left(\theta_{0},\theta_{1},\theta_{2},\theta_{3}\right)\in\mathbb{R}^{4}$
the estimable parameters of interest.\footnote{Further conditioning covariates can of course be included (for example
location or other fixed effects, which we will do in some specifications
considered in section \ref{subsec:Instrumental-treatment-effect}),
but for brevity we will omit them in equation \eqref{eq:BeliefGLM}.} The parameters $\left(\theta_{0},\theta_{2}\right)$ can be interpreted
as representing \emph{systematic} inflow of information that is not
controlled by treatment, and the error in the latent form of the linear
predictor, denoted $e$ for later reference, as representing \emph{idiosyncratic}
inflow of non-controlled information. 

The parameters $\left(\theta_{1},\theta_{3}\right)$ are pivotal for
the treatment effect. Specifically, by randomization $z$ will be
statistically independent from the potential beliefs $\left(b_{0},b_{1}\right)$,
such that by \eqref{eq:ObsBelief} 

\[
\mathrm{E}\left(b\mid c,z\right)=\mathrm{E}\left(b_{z}\mid c,z\right)=\mathrm{E}\left(b_{z}\mid c\right)
\]
and in turn, using the definition of $\Delta b$ and the GLM specification
\eqref{eq:BeliefGLM},
\begin{eqnarray*}
\mathrm{E}\left(b_{1}-b_{0}\mid c\right) & = & \mathrm{E}\left(b\mid c,z=1\right)-\mathrm{E}\left(b\mid c,z=0\right)\\
 & = & \mathrm{E}\left(\Delta b\mid c,z=1\right)-\mathrm{E}\left(\Delta b\mid c,z=0\right)\\
 & = & \begin{cases}
L\left(\theta_{0}+\theta_{1}\right)-L\left(\theta_{0}\right) & c=0\\
L\left(\theta_{0}+\theta_{1}+\theta_{2}+\theta_{3}\right)-L\left(\theta_{0}+\theta_{2}\right) & c=1
\end{cases}
\end{eqnarray*}
The left-hand side is the class-specific average treatment effect
(ATE), which is by the last equation identified by the predictive
margin difference of the fitted model. The monotonicity assumption
about $\psi$ implies the hypothesis that the ATE is positive in the
``below-group'' ($c=0$) and negative in the ``above-group'' ($c=1$).\footnote{The former is true if and only if $\theta_{1}>0$, and the latter
if and only if $\theta_{1}+\theta_{3}<0$, which jointly implies $\theta_{3}<-\theta_{1}<0$.} It can be tested by Wald tests applied to the ATE estimates and the
delta-method standard errors of estimates, which we will do in Section
\ref{subsec:Instrumental-treatment-effect}. 

The GLM model \eqref{eq:BeliefGLM} is generally estimable by maximum
likelihood estimation methods (MLE, or alternatively Bayesian methods).
However, since the results feed as inputs into the estimation of the
participation model, due care is needed with respect to specification.
We devote attention to two potentially important issues in section
\ref{subsec:Instrumental-treatment-effect}. First, our sample is
clustered by location (a subject's home city, there are four possible
clusters), date of study enrollment (the date a subject has participated
in the first survey, there are six possible days), and date of treatment
(the date a subject has participated in the second survey, there are
five possible days), and each cluster may have specific effects. We
will check for this by means of mixed-effects modeling, allowing for
crossed random effects in the three clustering-dimensions. Second,
the support of the dependent variable $\Delta b$ is limited to the
interval $\left[-1,1\right]$. We will check whether estimates from
non-linear specifications that take account of this fact differ significantly
from linear model estimates. Throughout, we will follow Occam's Razor.
It turns out that a simple linear model provides decent results.

\subsection{Augmented participation model\label{subsec:Augmented}}

We augment the participation model \eqref{eq:ProbitModel} in two
ways. First, note that it can be expressed equivalently as a function
of the \emph{change} of belief $\Delta b$, as defined above, 
\begin{equation}
\mathrm{Pr}\left(a=1\mid\Delta b,\boldsymbol{x}\right)=\Phi\left(\alpha+\beta\cdot\Delta b+\boldsymbol{x}\cdot\boldsymbol{\gamma}\right)\label{eq:ProbitModelAugm}
\end{equation}
This changes the interpretation of parameter $\alpha$ and of the
predictive margins slightly (the average participation probability
if everybody would have the given value of $\Delta b$, averaged over
$\boldsymbol{x}$), but it is easy to see that the interpretation
of $\beta$ and (since $\partial\Delta b/\partial b=1$) the marginal
effect of beliefs is substantively preserved,
\[
\frac{\partial\mathrm{Pr}\left(a=1\mid\Delta b,\boldsymbol{x}\right)}{\partial\Delta b}=\frac{\partial\mathrm{Pr}\left(a=1\mid\Delta b,\boldsymbol{x}\right)}{\partial b}=\varphi\left(\alpha+\beta\cdot\Delta b+\boldsymbol{x}\cdot\boldsymbol{\gamma}\right)\cdot\beta
\]
such that the meaning of hypothesis $\mathrm{H}_{0}:\beta=0$ is still
the same.\footnote{To be precise, we will report the APE based on the average structural
function \citep{Blundell.2004}. There are other approaches \citep{Lewbell.2012},
but the average structural function approach has decided advantages
\citep{Lin.2015}.} Yet, \eqref{eq:ProbitModelAugm} has the advantages described above.\footnote{Specifically, Kendall's rank correlation coefficient between $a$
and $b$ is $.0884$, and statistically significantly different from
zero (tie-corrected Kendall's score $43279\pm10576.1,$ continuity
corrected $p=.0000$). Thus, this correlation has the \emph{opposite}
direction of the causal effect that we uncover in Section \ref{sec:Results}.} 

Second, using the belief updating model \eqref{eq:BeliefGLM} in conjunction
with \eqref{eq:ProbitModelAugm}, a control function approach can
be used in which the residuals $\hat{e}$ from the belief updating
model are employed to fit
\begin{equation}
\mathrm{Pr}\left(a=1\mid\Delta b,\boldsymbol{x},\hat{e}\right)=\Phi\left(\alpha+\beta\cdot\Delta b+\eta\cdot\hat{e}+\boldsymbol{x}\cdot\boldsymbol{\gamma}\right)\label{eq:ProbitModelAugm2}
\end{equation}
by MLE, where $\eta\in\mathbb{R}$ is a fixed parameter \citep{Rivers.1988,Blundell.1989}.\footnote{In principle, parameter $\eta$ is estimable and can be used to test
for the endogeneity of beliefs, but since MLE estimates the parameters
in the belief updating and participation models jointly, it is not
actually estimated. Instead a test for zero correlation between the
residuals checks for endogeneity (reported in Section \ref{sec:Results}).} We will report results including location and survey date fixed effects,
and with and without bootstrap standard errors that account for clustering
by location, date of study enrollment (i .e. date of first survey
done), and date of treatment (i. e. date of second survey done). In
the Appendix we will also report estimates from a traditional two-step
approach in which the belief updating model predictions $\widehat{\Delta b}$
are used to fit 
\begin{equation}
\mathrm{Pr}\left(a=1\mid\widehat{\Delta b},\boldsymbol{x}\right)=\Phi\left(\alpha+\beta\cdot\widehat{\Delta b}+\boldsymbol{x}\cdot\boldsymbol{\gamma}\right)\label{eq:ProbitModelAugm3}
\end{equation}
by Newey's efficient minimum $\chi^{2}$ method \citep{Newey.1987}.
A key advantage of the control function method is that it estimates
the parameters and their variances separately, whereas Newey\textquoteright s
estimator yield variance-normalized estimates that are cumbersome
to interpret and which cannot directly be compared to the MLE estimates
\citep[see][ pp. 585-594, for a detailed discussion]{Wooldridge.2010}.

\subsection{Test of hypothesis and soundness of estimates\label{subsec:Test-of-hypothesis}}

We directly test the null hypothesis $\mathrm{H}_{0}:\beta=0$ against
the alternative $\mathrm{H}_{1}:\beta\neq0$ by a single-parameter
Wald test. The square-root of the observed Wald statistic ($\sqrt{W})$
is under the null hypothesis equal to the ratio of the MLE estimate
$\hat{\beta}$ and its standard error, and follows an asymptotic normal
($z$) distribution \citep[p. 89]{Davidson.1993}. We consider the
null hypothesis rejected if the observed $\sqrt{W}$ is outside the
critical region implied by a false-positive probability threshold
(level of significance) of five percent. We will report the $p$-value
for easy evaluation under different significance thresholds.

The APE estimate can also be understood within the potential outcomes
framework as a so-called local average treatment effect \citep[LATE,][]{Imbens.1994,Angrist.1996}.
This perspective clarifies that it hinges on two critical assumptions
that are fundamentally untestable (because each subject is only observed
under one of the two experimental conditions): the independence or
valid instrument assumption, and the monotonicity assumption. In our
setting the monotonicity assumption means that beliefs follow the
direction of the informational stimulus, as formalized by \eqref{eq:BeliefTheory},
and the valid instrument assumption means that any effect of treatment
on participation is fully mediated by beliefs.\footnote{Technically, the latter assumption requires all potential outcomes
(potential beliefs and potential participation choices in our setting)
to be statistically independent from treatment status, and that treatment
has some effect on beliefs in expectation. That this essentially amounts
to an exclusion restriction is readily apparent from a causal mediation
framework perspective \citep{MacKinnon.2007,Imai.2010,Pearl.2014,Preacher.2015}:
treatment is assumed to have a direct effect on beliefs but no (independent)
direct effect on participation, such that the total effect of treatment
on participation is purely indirect. } \citet{Mourifie.2017} derived testable implications of the two assumptions
in the form of two conditional moment inequalities, which can be tested
by a conditional likelihood ratio test in the intersection bounds
framework \citep{Chernozhukov.2013}. We will apply the local method
of the test to each of the ``above-'' and ``below-groups'' separately,
because the monotonicity assumption goes in opposite directions. We
consider our estimates as ``sound'', in the sense of being consistent
with the valid instrument and monotonicity assumptions, if none of
the tests rejects the null hypothesis that the two conditional moment
inequalities are consistent with the data at a significance threshold
of no less than ten percent.

\section{Results\label{sec:Results}}

\subsection{Instrumental effect of treatment on beliefs\label{subsec:Instrumental-treatment-effect}}

\begin{figure}[t]
\caption{Kernel density plots of change in belief ($\Delta b$) split by experimental
condition in the ``below'' group (a) and the ``above'' group (b).\label{fig:Kernels}}

\centering{}\subfloat[]{\includegraphics[width=0.5\textwidth]{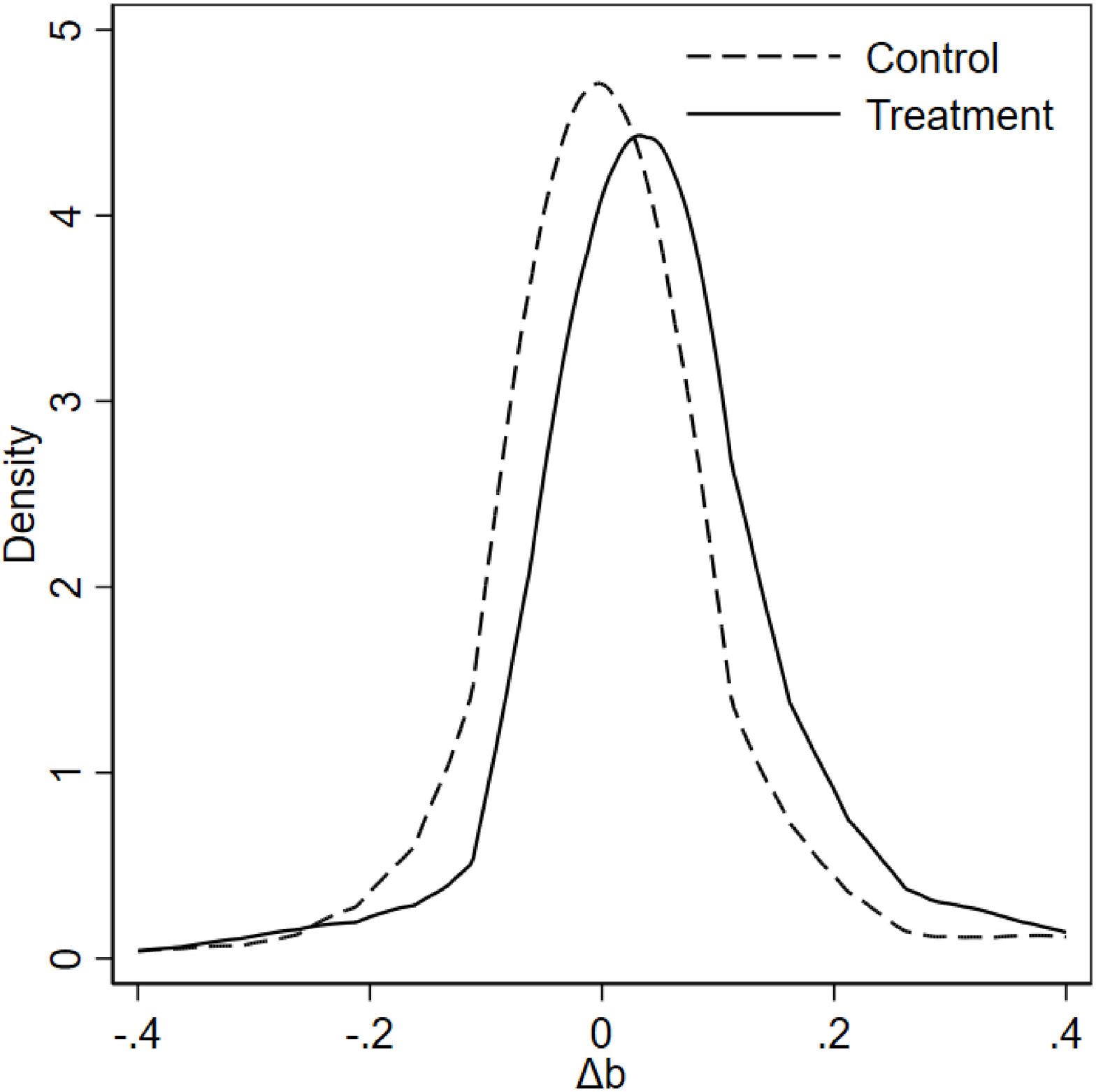}

}\subfloat[]{\includegraphics[width=0.5\textwidth]{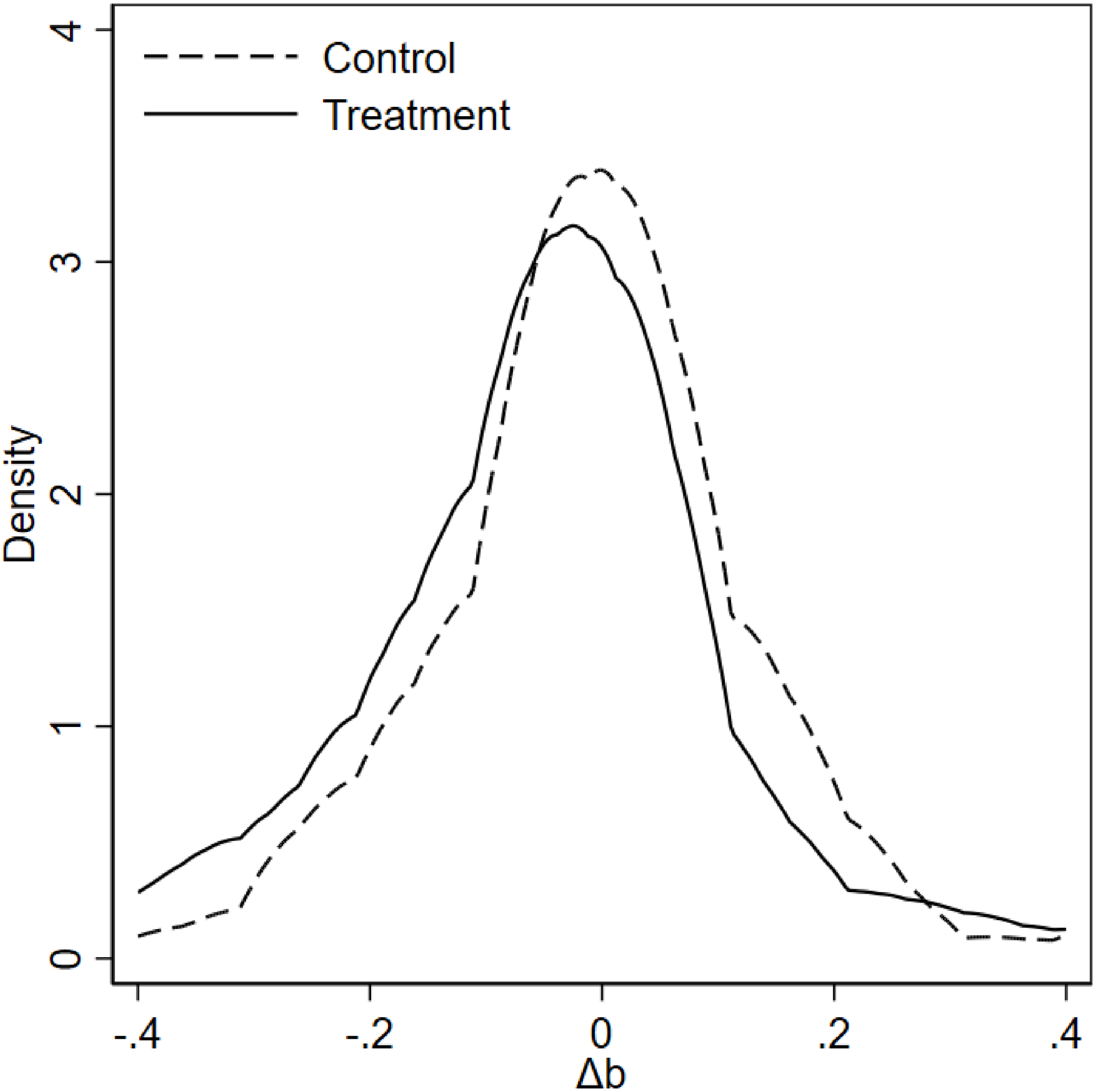}

}\textit{\footnotesize{}\medskip{}
}\\
\textit{\scriptsize{}Figure notes:}{\scriptsize{} The plots use an
Epanechnikov (parabolic) kernel with a 0.05 smoothing parameter. Estimation
is restricted to the interval $\left[-0.4,0.4\right]$ for brevity,
which includes masses $.9838$ in the ``below'' group and $.9471$
in the ``above'' group.}{\scriptsize\par}
\end{figure}
Figure \ref{fig:Kernels} visualizes the effect of treatment on beliefs
by kernel density plots of observed $\Delta b$ split by experimental
condition. As expected, the distribution under treatment is skewed
to the right in the ``below'' group (panel a) and to the left in
the ``above'' group (panel b) relative to control, respectively.

\begin{table}[t]
\caption{Average treatment effect (ATE) estimates and standard errors of estimates
(SEE) based on a linear specification of the belief updating model
\eqref{eq:BeliefGLM}.\label{tab:BeliefsATEs_linear}}

\begin{centering}
\begin{tabular*}{1\textwidth}{@{\extracolsep{\fill}}lrrrrr}
\toprule 
 & \multicolumn{2}{c}{Crossed random effects REML} &  & \multicolumn{2}{c}{Fixed effects OLS}\tabularnewline
\cmidrule{2-3} \cmidrule{3-3} \cmidrule{5-6} \cmidrule{6-6} 
ATE & Estimate & Delta-SEE ($p$) &  & Estimate & Delta-SEE ($p$)\tabularnewline
\midrule
Below ($c=0$) & $.0425$ & $.0097$ $\left(.000\right)$ &  & $.0425$ & $.0097$ $\left(.000\right)$\tabularnewline
Above ($c=1$) & $-.0555$ & $.0123$ $\left(.000\right)$ &  & $-.0554$ & $.0123$ $\left(.000\right)$\tabularnewline
\bottomrule
\end{tabular*}
\par\end{centering}
\centering{}\textit{\footnotesize{}\medskip{}
}\\
\textit{\scriptsize{}Table notes:}{\scriptsize{} The estimates and
SEEs are based on the regressions reported in tables \ref{tab:BeliefsREML}
and \ref{tab:BeliefsOLS} in the Appendix. SEEs are derived from the
regressions by the delta-method. The $p$-value of a Wald test of
the null hypothesis that the ATE estimate is equal to zero is reported
in parentheses. Rejections are ``highly significant'' for $p<.01$,
``significant'' for $p<.05$, and ``marginally significant'' for
$p<.1$.}{\scriptsize\par}
\end{table}
ATE estimates are listed in Tables \ref{tab:BeliefsATEs_linear} and
\ref{tab:BeliefsATEs_nonlinear}. The left-hand panel of Table \ref{tab:BeliefsATEs_linear}
(``Crossed random effects REML'') shows the estimates and delta-method
SEEs based on the mixed-effects restricted maximum likelihood regression
allowing for crossed random effects of location, enrollment date and
treatment date with unstructured variances and covariances.\footnote{Full estimation results are shown in Table \ref{tab:BeliefsREML}
in the Appendix. It also includes the results of a variant that allows
for random coefficients at the location level, but like for the random
intercepts the respective variances are all close to zero.} The estimated random effects except the residual are all close to
zero, and a likelihood-ratio test rejects the model against a simple
fixed effects model ($p=.9997$). Thus, by Occam's Razor, the latter
is good enough. Respective ATE estimates and delta-method SEEs are
shown in the right-hand panel (``Fixed effects OLS'') of table \ref{tab:BeliefsATEs_linear}.\footnote{Full estimation results are shown in Table \ref{tab:BeliefsOLS} in
the Appendix. It also reports bootstrap and cluster-bootstrap SEEs
with location, enrollment date and treatment date, respectively, defining
clusters. Differences to the pooled SEEs are negligible. A specification
including location or date fixed effects does produce almost identical
results, as all location or date fixed effects are not significantly
different from zero. For brevity, those estimates are omitted, but
they can be easily replicated with the online materials provided on
OSF \citep{JarkeNeuert.2021mat}.} Evidently, the results are almost identical to the left-hand panel.

The constant ($\theta_{0}$) is estimated to be not significantly
different from zero, such that there is no systematic change of beliefs
between the first survey (pre-intervention, $b^{\prime}$) and the
second survey (post-intervention, $b$) under control (namely $.243\pm.009$
vs. $.250\pm.009$ on average). As expected, treatment causes an upwards
adjustment of belief in the average subject belonging to the ``below-group'',
and a downwards adjustment in the ``above-group''. The magnitudes
are 4.2 and 5.5 percentage-points, respectively, and they are clearly
different from zero under all conventional significance levels. 

\begin{table}
\caption{Average treatment effect (ATE) estimates and standard errors of estimates
(SEE) based on non-linear specifications of the belief updating model
\eqref{eq:BeliefGLM} with transformation $2^{-1}\cdot\left(\Delta b+1\right)$.\label{tab:BeliefsATEs_nonlinear}}

\begin{centering}
\begin{tabular*}{1\textwidth}{@{\extracolsep{\fill}}lrrrrr}
\toprule 
 & \multicolumn{2}{c}{Fractional probit} &  & \multicolumn{2}{c}{Beta probit}\tabularnewline
\cmidrule{2-3} \cmidrule{3-3} \cmidrule{5-6} \cmidrule{6-6} 
ATE & Estimate & Delta-SEE ($p$) &  & Estimate & Delta-SEE ($p$)\tabularnewline
\midrule
Below ($c=0$) & $.0212$ & $.0037$ $\left(.000\right)$ &  & $.0212$ & $.0050$ $\left(.000\right)$\tabularnewline
Above ($c=1$) & $-.0277$ & $.0072$ $\left(.000\right)$ &  & $-.0286$ & $.0964$ $\left(.000\right)$\tabularnewline
\bottomrule
\end{tabular*}
\par\end{centering}
\centering{}\textit{\footnotesize{}\medskip{}
}\\
\textit{\scriptsize{}Table notes:}{\scriptsize{} The estimates and
SEEs are based on the regressions reported in tables \ref{tab:BeliefsFP}
and \ref{tab:BeliefsBeta} in the Appendix. SEEs are derived from
the regressions by the delta-method. The $p$-value of a Wald test
of the null hypothesis that the ATE estimate is equal to zero is reported
in parentheses. Rejections are ``highly significant'' for $p<.01$,
``significant'' for $p<.05$, and ``marginally significant'' for
$p<.1$.}{\scriptsize\par}
\end{table}
Those magnitudes are also robust to non-linear specifications of the
belief updating model. Specifically, the ATE estimates shown in table
\ref{tab:BeliefsATEs_nonlinear} take account of the fact that belief
changes are restricted to the interval $\left[-1,1\right]$. They
are based on fractional probit and beta probit regressions with transformation
$2^{-1}\cdot\left(\Delta b+1\right)$.\footnote{The full estimation results are shown in Tables \ref{tab:BeliefsFP}
and \ref{tab:BeliefsBeta} in the Appendix. Again, they also report
bootstrap-clustered SEEs with location, enrollment date (date of survey
1) and treatment date (date of survey 2), respectively, defining clusters.
Differences to the pooled SEEs are still negligible.} The re-transformed estimates are almost identical to the OLS estimates:
$.0424$ for the ``below-group'' and $-.0554$ (fractional probit)
resp. $-.0572$ (beta probit) in the ``above-group''. Again by the
principle of parsimony the OLS estimates are good enough.

In sum, the intervention was successful and treatment status is a
strong instrument for beliefs in the protest participation model.

\subsection{Average partial effect of a change of belief on participation}

\begin{table}[t]
\caption{Predictive margins with standard errors of estimates (SEE) based on
maximum likelihood estimation of the augmented probit participation
model \eqref{eq:ProbitModelAugm2}\label{tab:PredMargs}}

\begin{centering}
\begin{tabular*}{1\textwidth}{@{\extracolsep{\fill}}lrrrr}
\toprule 
 & \multicolumn{1}{c}{} & \multicolumn{3}{c}{Delta-SEE ($p$-value)}\tabularnewline
\cmidrule{3-5} \cmidrule{4-5} \cmidrule{5-5} 
 & Estimate & Standard & Bootstrap & Clustered bootstrap\tabularnewline
\midrule
Overall & $.1099$ & $.0080$ $\left(.000\right)$ & $.0081$ $\left(.000\right)$ & $.0087$ $\left(.000\right)$\tabularnewline
At means & $.0485$ & $.0143$ $\left(.001\right)$ & $.0156$ $\left(.002\right)$ & $.0138$ $\left(.000\right)$\tabularnewline
\midrule
At $\Delta b=-.4$ & $.5894$ & $.1541$ $\left(.000\right)$ & $.1620$ $\left(.000\right)$ & $.1436$ $\left(.000\right)$\tabularnewline
At $\Delta b=-.2$ & $.3308$ & $.0812$ $\left(.000\right)$ & $.0849$ $\left(.000\right)$ & $.0755$ $\left(.000\right)$\tabularnewline
At $\Delta b=0$ & $.1376$ & $.0151$ $\left(.000\right)$ & $.0155$ $\left(.000\right)$ & $.0151$ $\left(.000\right)$\tabularnewline
At $\Delta b=.2$ & $.0425$ & $.0083$ $\left(.000\right)$ & $.0089$ $\left(.000\right)$ & $.0087$ $\left(.000\right)$\tabularnewline
At $\Delta b=.4$ & $.0103$ & $.0048$ $\left(.032\right)$ & $.0052$ $\left(.045\right)$ & $.0048$ $\left(.032\right)$\tabularnewline
\bottomrule
\end{tabular*}
\par\end{centering}
\centering{}\textit{\footnotesize{}\medskip{}
}\\
\textit{\scriptsize{}Table notes:}{\scriptsize{} The estimates and
SEEs are based on the regressions reported in Table \ref{tab:ParticipationCF}.
SEEs are derived from the regressions by the delta-method. Bootstrap
SEE are based on 1,000 replications, clustered bootstrap SEE allow
for correlation of errors within strata formed by crossing locations,
enrollment dates, and treatment dates (86 populated clusters). The
$p$-value of a Wald test of the null hypothesis that the respective
margin estimate is equal to zero is reported in parentheses. Rejections
are ``highly significant'' for $p<.01$, ``significant'' for $p<.05$,
and ``marginally significant'' for $p<.1$. The mean of $\Delta b$
is equal to $.0095397$.}{\scriptsize\par}
\end{table}
\begin{figure}
\caption{Predictive margins with 95\% confidence intervals at different values
of $\Delta b$ based on maximum likelihood estimation of the augmented
probit participation model \eqref{eq:ProbitModelAugm2}\label{fig:PredMarg}}

\centering{}\includegraphics[width=0.75\textwidth]{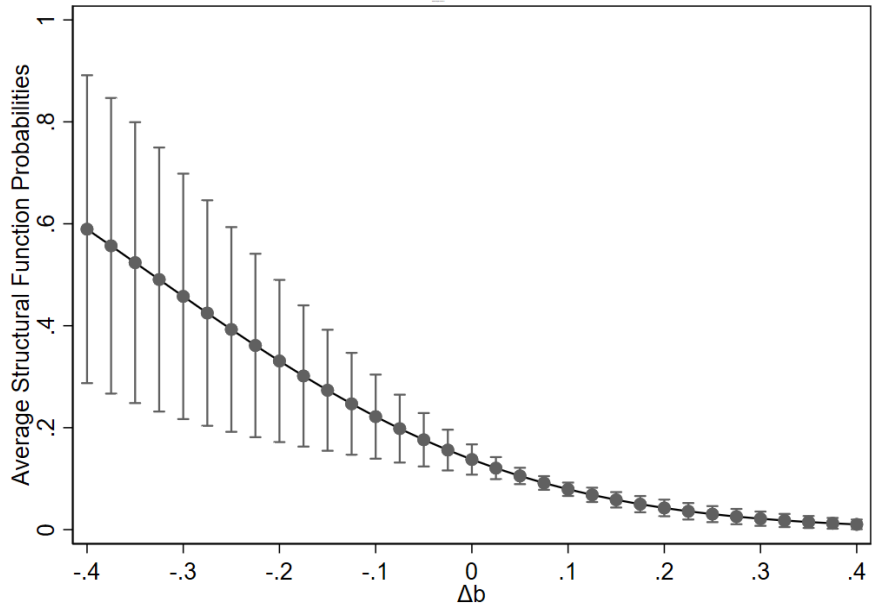}
\end{figure}
The fitted participation model \eqref{eq:ProbitModelAugm2} predicts
the observed share of participants in the sample (Table \ref{tab:SumPart})
very accurately, as evident from the overall predictive margin estimate
in the first row of Table \ref{tab:PredMargs}, which is based on
the MLE estimation of the augmented participation model \eqref{eq:ProbitModelAugm2}.\footnote{The full estimation results are shown in Table \ref{tab:ParticipationCF}
in the Appendix. At the local level, the empirical participation share
versus specific predictive margin comparisons are $.0918$ versus
$.0909$ in Berlin, $.1128$ versus $.1126$ in Hamburg, $.1036$
versus $.1044$ in Munich, and $.1378$ versus $.1386$ in Cologne.
In the Appendix there are also results obtained via Newey's efficient
minimum $\chi^{2}$ two-step method in Table \ref{tab:ParticipationTS}.
They are almost equal to the MLE estimates, but recall that they are
not directly comparable (see section \ref{sec:Methods}).} Table \ref{tab:PredMargs} also shows the margin at means and the
margins at selected fixed values of belief changes ($\Delta b$).
A finer resolution is illustrated in figure \ref{fig:PredMarg}. A
negative relationship between beliefs and the probability of participation
is clearly evident.

\begin{table}[t]
\caption{Average partial effect (APE) estimates with standard errors of estimates
(SEE) based on maximum likelihood estimation of the augmented probit
participation model \eqref{eq:ProbitModelAugm2}\label{tab:APEs}}

\begin{centering}
\begin{tabular*}{1\textwidth}{@{\extracolsep{\fill}}lrrrr}
\toprule 
 & \multicolumn{1}{c}{} & \multicolumn{3}{c}{Delta-SEE ($p$-value)}\tabularnewline
\cmidrule{3-5} \cmidrule{4-5} \cmidrule{5-5} 
 & Estimate & Standard & Bootstrap & Clustered bootstrap\tabularnewline
\midrule
Overall & $-.6787$ & $.2058$ $\left(.001\right)$ & $.2152$ $\left(.002\right)$ & $.1918$ $\left(.000\right)$\tabularnewline
At means & $-.3729$ & $.0501$ $\left(.000\right)$ & $.0501$ $\left(.000\right)$ & $.0478$ $\left(.000\right)$\tabularnewline
At pre. & $-.7090$ & $.2147$ $\left(.001\right)$ & $.2250$ $\left(.002\right)$ & $.1999$ $\left(.000\right)$\tabularnewline
At post. & $-.6841$ & $.2017$ $\left(.001\right)$ & $.2113$ $\left(.001\right)$ & $.1877$ $\left(.000\right)$\tabularnewline
\bottomrule
\end{tabular*}
\par\end{centering}
\centering{}\textit{\footnotesize{}\medskip{}
}\\
\textit{\scriptsize{}Table notes:}{\scriptsize{} The estimates and
SEEs are based on the regressions reported in Table \ref{tab:ParticipationCF}.
SEEs are derived from the regressions by the delta-method. Bootstrap
SEE are based on 1,000 replications, clustered bootstrap SEE allow
for correlation of errors within strata formed by crossing locations,
enrollment dates, and treatment dates (86 populated clusters). The
$p$-value of a Wald test of the null hypothesis that the respective
ATE estimate is equal to zero is reported in parentheses. Rejections
are ``highly significant'' for $p<.01$, ``significant'' for $p<.05$,
and ``marginally significant'' for $p<.1$. ``At pre.'' in the
third row means the APE evaluated at $\Delta b=0$ (which assumes
that all subjects have that value, ceteris paribus), which implies
that the post-intervention belief is equal to the pre-intervention
belief. Likewise, ``At post.'' in the bottom row means that the
APE evaluated at the observed mean of $\Delta b$ (namely $.0095397$).}{\scriptsize\par}
\end{table}
The value of the focal parameter $\beta$ is estimated $-3.3062$
and with standard errors robustly at around a quarter of that value,
such that the Wald test rejects $\mathrm{H}_{0}:\beta=0$ at all conventional
levels of significance ($p=.000$). It can be concluded that participation
behavior displays strategic substitutability. The APE estimates in
Table \ref{tab:APEs} quantify the magnitude. From the top row of
Table \ref{tab:APEs} we see that overall a one percentage-point increase
of belief causes a .67 percentage-point decrease in the probability
of participation in the average subject.\footnote{There is some spatial heterogeneity. The specific overall APE estimates
are $-.5928$ in Berlin, $-.6944$ in Hamburg, $-.6617$ in Munich,
and $-.7978$ in Cologne.} The remaining rows also provide estimates of the APE evaluated at
the means of all variables, and evaluated at $\Delta b=0$ (i. e.
if all subjects would have a post-intervention belief equal to their
pre-intervention belief, ``at pre.'') and at $\Delta b=.0095$,
the observed mean (``at post.'').

Finally, we turn to the ``soundness'' of the results as defined
in Section \ref{subsec:Test-of-hypothesis}. To begin with, a Wald
test against the null hypothesis that $\Delta b$ is exogenous in
the standalone model \eqref{eq:ProbitModelAugm} is rejected at all
conventional levels of significance with $\chi^{2}\left(1\right)=11.62$
and $p=.0007$. Thus, experimental control was in fact crucial for
recovering the causal effect of a change in belief on participation.
Finally, the conditional likelihood ratio intersection bound tests
following the Mourifié-Wan procedure do not reject the joint hypothesis
of instrument validity and monotonicity in any of the classes at $p<.1$.
We therefore conclude that our results are internally ``sound'',
in the sense that the estimated APE most likely indentifies the causal
LATE on the class of subjects that respond to treatment.

\section{Discussion\label{sec:Discussion}}

This paper connects to several bodies of literature spanning several
disciplines. First, it adds to the emerging literature on the structural,
tactical, and most importantly (for our purposes) communicational
properties of the climate protest movement. Evidence from this line
of research shows that the movement is broadly characterized by interpersonally
mobilized young females from well-educated backgrounds \citep{Wahlstrom.2019,deMoor.2020,Hayes.2021}
emphasizing science, peaceful resistance, social and political change,
sustainable lifestyle, and inter-generational justice \citep{Holmberg.2019,Marquardt.2020,Bugden.2020,Zabern.2021,Huttunen.2021},
but there seem to be subtle regional differences depending on prevailing
political and communicative institutions \citep{Kern.2021}. Specifically,
social media communication seems to emphasize group cohesion and emotional
attachment in some instances \citep{Segerberg.2011,Brunker.2019}
and functional information exchange (e. g. documentation and coordination
of events, protest tactics, transportation, turnout, police presence,
violence, medical services, legal support, etc.) in others \citep{Boulianne.2020}.\footnote{Notably, recent evidence suggests that COVID-19 left a footprint in
the communication structure by shifting the relative emphasis given
to non-functional information \citep{Hassler.2021}.} This is consistent with evidence from several other protest movements
summarized in \citet{Jost.2018}, showing that social media platforms
facilitate the exchange of information that is vital to both coordination
of protest activities and emotional and motivational contents. This
is an interesting complement to our study, as it illustrates the specific
mechanisms through which beliefs about turnout are shaped and correlated
``in the wild'' (i .e. without treatment intervention). That digital
communication indeed drives protest dynamics is demonstrated by \citet{Enikolopov.2020},
who used an instrumental variable approach to recover a causal effect
of the penetration of the dominant online social network (VK) on protest
activity in 2011 Russia.

Second, we also contribute to a recent stream of literature investigating
the motivational (preferences and beliefs) structure underlying climate
protest participation. Climate protesters tend to be instrumentally
(i. e. to attain a change of public policy) motivated \citep{deMoor.2020,Cologna.2021},
although there is regional and individual heterogeneity about advocated
means \citep{Beer.2020,Svensson.2021,Huttunen.2021b,Soliev.2021},
but there is also a strong affective-emotional basis, revolving around
feelings of worry, anxiety, frustration, and anger \citep{Wahlstrom.2019,deMoor.2020},
and a perceived moral duty to act \citep{Fernandes.2020,Wallis.2021}.
Social expression and self-signaling, identification, and event enjoyment
also play a role \citep{Walgrave.2012,Fernandes.2020,Wallis.2021,Cologna.2021}.\footnote{Evidence from two different protest movements in Spain collected by
\citet{Sabucedo.2017} suggests that concerns of justice are important
be for protestors. The inter-generational justice component in climate
action suggests that this might also important for climate protesters,
at least the young cohorts, but direct evidence on this is lacking. } This stream of literature supports our conclusion that both interactive
and idiosyncratic factors drive participation decisions. We also believe
that the beliefs-preferences-constraints terminology (providing an
interface to both decision theoretic analysis and behavioral research)
could serve as a powerful framework for organizing extant and further
research in this area \citep[see][for a general discussion]{Gintis.2005,Gintis.2014}.

Third, our study also contributes to the rapidly growing empirical
literature on protest and social movements more generally, such as
the recent social media driven ``Occupy'' movement \citep{Theocharis.2015},
``Idle No More'' \citep{Raynauld.2017}, the ``Tea Party'' movement
\citep{Madestam.2013}, ``Black Lives Matter'' \citep{Freelon.2018},
``March for Dignity'' \citep{Sabucedo.2017}, the ``Arab Spring''
\citep{Acemoglu.2018}, ``EuroMaidan'' \citep{Metzger.2017}, or
various youth movements \citep[e. g.][]{Theocharis.2012,Raynauld.2016},
but also historical cases such as the social movement against slavery
in the 19th century US \citep{Dippel.2021}. The literature is vast
and diverse, a review is beyond the scope of this section \citep[we refer to][]{Jost.2018}.
But we want to extend on three recent studies that are proximately
relevant for us, as they also bear on the strategic component in participation
decisions. 

The finding of strategic substitutability is consistent with a study
very similar to ours in the context of Hong Kong's Umbrella Movement
\citep{Cantoni.2019}. They used essentially the same informational
intervention to induce exogenous variation in survey-measured beliefs
within a student sample of potential protesters and also found that
an increase of belief concerning others' attendance in a protest reduces
the probability of participation. We do not only find similar results
in a different protest setting, but extend on their study by providing
more statistical power, a heterogenous sample, and a substantially
refined econometric framework. The facts that both studies use randomization
and control and results go in the same direction can be considered
solid evidence in favor of strategic substitutability .

But there is also compelling evidence in favor of strategic complementarity.
\citet{Manacorda.2020} investigate the role of digital information
and communication technology in mass political mobilization, using
georeferenced data on the coverage of mobile phone signal together
with data on protest incidences and individual participation decisions
for the entire African continent between 1998 and 2012. They find
that mobile phones are instrumental to mass mobilization (only) during
economic downturns (when reasons for grievance emerge and the cost
of participation falls) in a way being consistent with a network model
with imperfect information and strategic complementarities with respect
to neighbors' participation. \citet{Gonzalez.2020} studied 2011 high-school
student protests for reform of educational institutions in Chile,
using administrative data of daily school attendance and an identification
strategy exploiting partially overlapping networks and within school
exposure to an inaugural college protest. He finds causal evidence
of complementarities in school skipping decisions within student networks
in national protest days. The mixed evidence suggests that the conditions
and determinants of the direction of strategic interdependence seems
to be a valuable avenue for further research.\footnote{In a within a theoretical model, \citet{Shadmehr.2021} shows that
when a protest movement's goal is ``modest'', free-riding concerns
dominate making their actions strategic substitutes, whereas when
the movement's goal is to overthrow the entire status quo, coordination
concerns dominate, and actions become strategic complements. \citet{JarkeNeuert.2021}
derives a similar result drawing on step-level public goods terminology.}

Fourth and finally, our study relates to theoretical work protest
movement dynamics, its interface with public policy, and social outcomes.
Several recent theoretical studies have furthered traditional analysis
of protests movements by considering preferences for fairness and
justice \citep[e. g.][]{Passarelli.2017} and strategic uncertainty
\citep[e. g.][]{BuenodeMesquita.2010,Shadmehr.2011,Edmond.2013,Shadmehr.2021}.
The parametric participation model developed in \citet{JarkeNeuert.2021}
and estimated in this paper can capture such aspects parsimoniously,
and connects seamlessly with quantal response equilibrium analysis,
specifically the class of political participation games \citep{Goeree.2005,Goeree.2016}.
The parameter estimates provide empirically plausible restrictions,
and allow for numerically computable simulations. For instance, maximum
likelihood fits show that the empirical distributions of beliefs measured
in the present study fit nicely to a theoretical beta distribution
(see the notes to Tables \ref{tab:SumPriors} and \ref{tab:SumPosteriors})
with shape parameter about one and scale parameter about three. Drawing
on the concept of random beliefs \citep{Sandroni.2004,Friedman.2005},
the (calibrated) distribution can be used as an empirically plausible
seed in game theoretic equilibrium models, or to create pseudo-random
draws to equip populations of automata with empirically plausible
initial beliefs in simulations.

With an empirically accurate but tractable model of protest dynamics,
we believe that a connection to the impressive formal framework recently
presented by \citet{Egorov.2017} could provide for a powerful ``integrated
assessment model'' for studying the interaction between public regulation,
corporate self-regulation and activism, within the realm of climate
policy or elsewhere.

\section{Conclusion\label{sec:Conclusion}}

We conclude with a cautious outlook regarding the future of climate
protest. The results from our and related studies are mixed news for
the movement, as they suggest that the pre-COVID-19 momentum was close
to maximum capacity. There is evidence suggesting that the emotional
basis of the movement shows signs of erosion \citep{deMoor.2020},
and that the COVID-19-related lockdowns may have harmed the momentum
persistently \citep{Hassler.2021}. Our results suggest that the climate
protest movement also not spills over easily from the youth to the
adult population, at least in the German sample we studied. Specifically,
in equilibrium any exogenous increase (or decrease) in unconditional
motivation to join (parameter $\alpha$ in the model) will be (partially)
offset by ``free-riding'' behavior \citep{JarkeNeuert.2021}.

Nonetheless, it is also possible that some ``critical juncture''
induces a major turnaround \citep{Capoccia.2016}. Regarding eventual
impact of climate protest on actual climate policy,\footnote{Generally, and on historical scale, the fate of a protest movement
depends on a complex interaction of conditions that are broadly classifiable
as institutions and chance \citep{Acemoglu.2012,Acemoglu.2016,Hall.2009}. } there is currently no solid evidence. But there are two interesting
studies that exemplarily highlight two possible channels and could
inspire research in this area. A direct \emph{political} channel is
illustrated by \citet{Madestam.2013}, who investigate the hen-and-egg
problem of whether protests cause political change, or whether they
are merely symptoms of underlying shifts in policy preferences. They
study the US Tea Party protests on April 15, 2009, exploiting variation
in rainfall on that day as an instrumental variable for turnout. They
show that good weather had significant consequences for the subsequent
local strength of the movement, increased public support for Tea Party
positions, and led to more Republican votes in the 2010 midterm elections.
Policy making was also affected, as incumbents responded to large
protests in their district by voting more conservatively in Congress.
Thus, protests can indeed affect policy making.

A more \emph{economic} channel is suggested by \citet{Acemoglu.2018}.
Using daily variation in the number of participants in street protests
that brought down Mubarak's government in Egypt, the authors document
that more intense protests are associated with lower stock market
valuations for firms connected to the group currently in power relative
to non-connected firms, but have no impact on the relative valuations
of firms connected to rival groups. This suggests that the protests
served as a partial check on political rent-seeking.
\bibliographystyle{apalike}
\bibliography{Climate_Protest}
\newpage{}

\appendix

\section*{Appendix}

\begin{table}[p]
\caption{Population sizes of Berlin, Hamburg, Munich, and Cologne by the end
of 2019, and breakdowns by gender and age.\label{tab:PopProperties}}

\centering{}{\small{}}%
\begin{tabular*}{1\textwidth}{@{\extracolsep{\fill}}lrrrrrrrr}
\toprule 
 & \multicolumn{2}{c}{Berlin} & \multicolumn{2}{c}{Hamburg} & \multicolumn{2}{c}{Munich} & \multicolumn{2}{c}{Cologne}\tabularnewline
\midrule
Female & $1,904,052$ & $\left(.5051\right)$ & $943,279$ & $\left(.5106\right)$ & $789,041$ & $\left(.5058\right)$ & $557,563$ & $\left(.5109\right)$\tabularnewline
Male & $1,865,443$ & $\left(.4949\right)$ & $903,974$ & $\left(.4894\right)$ & $771,001$ & $\left(.4942\right)$ & $534,256$ & $\left(.4895\right)$\tabularnewline
\midrule
Age $<18$ & $605,098$ & $\left(.1605\right)$ & $310,886$ & $\left(.1683\right)$ & $236,921$ & $\left(.1519\right)$ & $176,088$ & $\left(.1613\right)$\tabularnewline
Age $\left[18,30\right[$ & $554,064$ & $\left(.1470\right)$ & $292,351$ & $\left(.1583\right)$ & $265,629$ & $\left(.1703\right)$ & $183,968$ & $\left(.1686\right)$\tabularnewline
Age $\left[30,50\right[$ & $1,135,428$ & $\left(.3012\right)$ & $542,869$ & $\left(.2939\right)$ & $499,694$ & $\left(.3203\right)$ & $323,154$ & $\left(.2961\right)$\tabularnewline
Age $\left[50,65\right[$ & $753,399$ & $\left(.1999\right)$ & $364,788$ & $\left(.1975\right)$ & $290,176$ & $\left(.1860\right)$ & $217,192$ & $\left(.1990\right)$\tabularnewline
Age $\geq65$ & $721,506$ & $\left(.1914\right)$ & $336,359$ & $\left(.1821\right)$ & $267,622$ & $\left(.1715\right)$ & $191,417$ & $\left(.1754\right)$\tabularnewline
\midrule
Total & $3,769,495$ &  & $1,847,253$ &  & $1,560,042$ &  & $1,091,819$ & \tabularnewline
\bottomrule
\end{tabular*}\textit{\footnotesize{}\medskip{}
}\textit{\scriptsize{}Table notes:}{\scriptsize{} Listed are the counts
of inhabitants in the respective municipality, broken down by gender
and age, with relative class sizes in parentheses. All data are taken
from official census records with reporting date December 31, 2019.
The raw data source is \citet{StatistikBBB.2019} for Berlin, \citet{StatistikNord.2020}
for Hamburg, \citet{StatistikMuc.2020,StatistikMuc.2020b} for Munich,
and \citet{StatistikCol.2020,StatistikCol.2020b,StatistikCol.2020c}
for Cologne. Own calculations performed where necessary to fit our
classification. Specifically for Cologne, a different classification
in the 18 up to 65 range is officially reported for the 2019 data
(classes 18 up to 25, 25 up to 45, and 45 up to 65), such that we
extrapolated the proportions corresponding to our classification (.254
for 18 up to 30, .446 for 30 up to 50, and .300 for 50 up to 65) from
end-of-2018 data \citep{StatistikCol.2020c} to the total of 724,314
in that class.}{\scriptsize\par}
\end{table}
\begin{table}[p]
\caption{Sample breakdowns by location, gender and age.\label{tab:SampleProperties}}

\centering{}%
\begin{tabular*}{1\textwidth}{@{\extracolsep{\fill}}lrrrrrrrrrr}
\toprule 
 & \multicolumn{2}{c}{Berlin} & \multicolumn{2}{c}{Hamburg} & \multicolumn{2}{c}{Munich} & \multicolumn{2}{c}{Cologne} & \multicolumn{2}{c}{Overall}\tabularnewline
\midrule
Female & $261$ & $\left(.5327\right)$ & $223$ & $\left(.5589\right)$ & $166$ & $\left(.5929\right)$ & $188$ & $\left(.5513\right)$ & $838$ & $\left(.5500\right)$\tabularnewline
Male & $227$ & $\left(.4633\right)$ & $174$ & $\left(.4361\right)$ & $114$ & $\left(.4071\right)$ & $151$ & $\left(.4428\right)$ & $666$ & $\left(.4411\right)$\tabularnewline
Diverse & $2$ & $\left(.0041\right)$ & $2$ & $\left(.0050\right)$ & $0$ & $\left(.0000\right)$ & $0$ & $\left(.0000\right)$ & $4$ & $\left(.0026\right)$\tabularnewline
No response & $0$ & $\left(.0000\right)$ & $0$ & $\left(.0000\right)$ & $0$ & $\left(.0000\right)$ & $2$ & $\left(.0059\right)$ & $2$ & $\left(.0013\right)$\tabularnewline
\midrule
Age less than 18 & $0$ & $\left(.0000\right)$ & $0$ & $\left(.0000\right)$ & $0$ & $\left(.0000\right)$ & $0$ & $\left(.0000\right)$ & $0$ & $\left(.0000\right)$\tabularnewline
Age 18 up to 30 & $85$ & $\left(.1735\right)$ & $83$ & $\left(.2080\right)$ & $65$ & $\left(.2321\right)$ & $71$ & $\left(.2082\right)$ & $304$ & $\left(.2013\right)$\tabularnewline
Age 30 up to 50 & $187$ & $\left(.3816\right)$ & $159$ & $\left(.3985\right)$ & $120$ & $\left(.4286\right)$ & $117$ & $\left(.3431\right)$ & $583$ & $\left(.3861\right)$\tabularnewline
Age 50 up to 65 & $160$ & $\left(.3265\right)$ & $124$ & $\left(.3108\right)$ & $76$ & $\left(.2714\right)$ & $128$ & $\left(.3754\right)$ & $488$ & $\left(.3232\right)$\tabularnewline
Age 65 and more & $58$ & $\left(.1184\right)$ & $33$ & $\left(.0827\right)$ & $19$ & $\left(.0679\right)$ & $25$ & $\left(.0734\right)$ & $135$ & $\left(.0894\right)$\tabularnewline
\midrule
Total & $490$ &  & $399$ &  & $280$ &  & $341$ &  & $1,510$ & \tabularnewline
\bottomrule
\end{tabular*}\textit{\footnotesize{}\medskip{}
}\textit{\scriptsize{}Table notes:}{\scriptsize{} Listed are the counts
of subjects that completed all three surveys in the respective location
class, broken down by self-resported gender and age, with relative
class sizes in perantheses. A Kruskal-Wallis equality-of-populations
rank test does not reject equality of distributions across locations
for gender ($\chi^{2}\left(3\right)=2.790$ with ties, $p=.4252$)
but for age at a five-percent level of significance ($\chi^{2}\left(3\right)=9.487$
with ties, $p=.0235$). This is due to Munich, where the mean ($\pm$
standard deviation) is lower at $42.42\pm14.24$, as compared to Berlin
at $45.72\pm14.39$, Hamburg at $44.14\pm14.44$, and Cologne at $44.54\pm14.75$.
This corresponds to the overall population properties shown in Table
\ref{tab:PopProperties}, where Munich has a slight bent towards younger
ages. }{\scriptsize\par}
\end{table}
\begin{table}[p]
\caption{Probability-expected sampling frequencies based on the gender and
age distributions in the local populations.\label{tab:Diagnostics}}

\centering{}%
\begin{tabular*}{1\textwidth}{@{\extracolsep{\fill}}lrrrrrrrr}
\toprule 
 & \multicolumn{2}{c}{Berlin} & \multicolumn{2}{c}{Hamburg} & \multicolumn{2}{c}{Munich} & \multicolumn{2}{c}{Cologne}\tabularnewline
\midrule
Female & $247.499$ & $\left(.2401\right)$ & $203.7294$ & $\left(.0057\right)$ & $141.624$ & $\left(.0040\right)$ & $174.2169$ & $\left(.1436\right)$\tabularnewline
\midrule
Age 18 up to 30 & $85.7956$ & $\left(1.000\right)$ & $75.9246$ & $\left(.3719\right)$ & $56.2126$ & $\left(.2043\right)$ & $68.5060$ & $\left(.7355\right)$\tabularnewline
Age 30 up to 50 & $175.8186$ & $\left(.3003\right)$ & $140.985$ & $\left(.0667\right)$ & $105.746$ & $\left(.0843\right)$ & $120.336$ & $\left(.7341\right)$\tabularnewline
Age 50 up to 65 & $116.6622$ & $\left(.0000\right)$ & $94.7368$ & $\left(.0008\right)$ & $61.4073$ & $\left(.0428\right)$ & $80.8780$ & $\left(.0000\right)$\tabularnewline
Age 65 and more & $111.7236$ & $\left(.0000\right)$ & $87.3536$ & $\left(.0000\right)$ & $56.6344$ & $\left(.0000\right)$ & $71.2799$ & $\left(.0000\right)$\tabularnewline
\bottomrule
\end{tabular*}\textit{\footnotesize{}\medskip{}
}\textit{\scriptsize{}Table notes:}{\scriptsize{} Listed are the expected
$k$'s based on the relative frequencies in the overall local populations
listed in Table \ref{tab:PopProperties}, and the $p$-values of two-sided
binomial probability tests of the null hypothesis that the observed
$k$ (Table \ref{tab:SampleProperties}) are equal to the expected
$k$, respectively, in parentheses. }{\scriptsize\par}
\end{table}

\begin{table}[p]
\caption{Restricted maximum likelihood estimates and standard errors of estimates
(SEE) of the linear belief updating model \eqref{eq:BeliefGLM} with
crossed random effects.\label{tab:BeliefsREML}}

\begin{centering}
\begin{tabular*}{1\textwidth}{@{\extracolsep{\fill}}lrrrrr}
\toprule 
 & \multicolumn{2}{r}{with location random slope} &  & \multicolumn{2}{c}{without location random slope}\tabularnewline
\cmidrule{2-3} \cmidrule{3-3} \cmidrule{5-6} \cmidrule{6-6} 
 & Estimate & SEE ($p$) &  & Estimate & SEE ($p$)\tabularnewline
\midrule
$\theta_{0}$ & $.0072$ & $.0080$ $\left(.371\right)$ &  & $.0072$ & $.0080$ $\left(.371\right)$\tabularnewline
$\theta_{1}$ & $.0425$ & $.0097$ $\left(.000\right)$ &  & $.0425$ & $.0097$ $\left(.000\right)$\tabularnewline
$\theta_{2}$ & $-.0003$ & $.0132$ $\left(.982\right)$ &  & $-.0003$ & $.0130$ $\left(.982\right)$\tabularnewline
$\theta_{3}$ & $-.0979$ & $.0159$ $\left(.000\right)$ &  & $-.0979$ & $.0157$ $\left(.000\right)$\tabularnewline
\midrule
Log rest. $\mathcal{L}$ &  & $835.05084$ &  &  & $835.04914$\tabularnewline
Wald $\chi^{2}\left(3\right)$ &  & $124.24$ &  &  & $124.64$\tabularnewline
Model $p$ &  & $.0000$ &  &  & $.0000$\tabularnewline
\midrule
LR test $\chi^{2}\left(12\right)$ resp. $\chi^{2}\left(3\right)$ &  & $.01$ &  &  & $.01$\tabularnewline
LR test $p$ &  & $1.0000$ &  &  & $.9997$\tabularnewline
\bottomrule
\end{tabular*}
\par\end{centering}
\centering{}\textit{\footnotesize{}\medskip{}
}\\
\textit{\scriptsize{}Table notes:}{\scriptsize{} Each regression has
1,510 observations. Random intercepts are included at the level of
location (four), enrollment date (six), and treatment date (five).
The regression in the left-hand panel (``with random slope'') also
includes random slopes at the location level, the right-hand panel
(``without random slope'') does not. Variances and covariances are
unstructured. All estimated variances and covariances are close to
zero ($>-.0001$ and $<.0001$) except the residual variances, which
are $.0190$ in both models. The two rows at the bottom report the
results of a likelihood-ratio test against a simple linear (fixed
effects) model. For the standard errors of estimates, the $p$-value
of a Wald test of the null hypothesis that the coefficient estimate
is equal to zero is reported in parentheses. The bottom row reports
the $p$-value of a Wald test against the null hypothesis that all
coefficient estimates are jointly zero. Rejections are ``highly significant''
for $p<.01$, ``significant'' for $p<.05$, and ``marginally significant''
for $p<.1$.}{\scriptsize\par}
\end{table}
\begin{table}[p]
\caption{Ordinary least squares estimates and standard errors of estimates
(SEE) of the linear belief updating model \eqref{eq:BeliefGLM}.\label{tab:BeliefsOLS}}

\begin{centering}
\begin{tabular*}{1\textwidth}{@{\extracolsep{\fill}}lrrrrr}
\toprule 
 & \multicolumn{1}{c}{} &  & \multicolumn{3}{c}{SEE ($p$-value)}\tabularnewline
\cmidrule{4-6} \cmidrule{5-6} \cmidrule{6-6} 
 & Estimate &  & Standard & Bootstrap & Cluster bootstrap\tabularnewline
\midrule
$\theta_{0}$ & $.0071$ &  & $.0079$ $\left(.369\right)$ & $.0058$ $\left(.218\right)$ & $.0048$ $\left(.137\right)$\tabularnewline
$\theta_{1}$ & $.0425$ &  & $.0097$ $\left(.000\right)$ & $.0074$ $\left(.000\right)$ & $.0057$ $\left(.000\right)$\tabularnewline
$\theta_{2}$ & $-.0003$ &  & $.0130$ $\left(.984\right)$ & $.0131$ $\left(.984\right)$ & $.0117$ $\left(.982\right)$\tabularnewline
$\theta_{3}$ & $-.0979$ &  & $.0157$ $\left(.000\right)$ & $.0166$ $\left(.000\right)$ & $.0151$ $\left(.000\right)$\tabularnewline
\midrule
$F\left(3,1506\right)$ &  &  & $41.50$  &  & \tabularnewline
Wald $\chi^{2}\left(3\right)$ &  &  &  & $112.36$ & $135.43$\tabularnewline
Model $p$ &  &  & $.0000$ & $.0000$ & $.0000$\tabularnewline
\bottomrule
\end{tabular*}
\par\end{centering}
\centering{}\textit{\footnotesize{}\medskip{}
}\\
\textit{\scriptsize{}Table notes:}{\scriptsize{} Each regression has
1,510 observations, an $R^{2}$ of $.0763$, an adjusted $R^{2}$
of $.0745$, and a root MSE of $.1379$. Bootstraps involve 1,000
replications each. There are $4\times6\times5=120$ feasible clusters
of which 86 are populated. A specification including location or date
fixed effects does produce almost identical results, as all location
or date fixed effects are not significantly different from zero. For
the standard errors of estimates, the $p$-value of a Wald test of
the null hypothesis that the coefficient estimate is equal to zero
is reported in parentheses. The bottom row reports the $p$-value
of a Wald test against the null hypothesis that all coefficient estimates
are jointly zero. Rejections are ``highly significant'' for $p<.01$,
``significant'' for $p<.05$, and ``marginally significant'' for
$p<.1$. }{\scriptsize\par}
\end{table}
\begin{table}[p]
\caption{Maximum pseudo-likelihood estimates and standard errors of estimates
(SEE) of the non-linear belief updating model \eqref{eq:BeliefGLM}
in fractional probit regression specification belief with transformation
$2^{-1}\cdot\left(\Delta b+1\right)$.\label{tab:BeliefsFP}}

\begin{centering}
\begin{tabular*}{1\textwidth}{@{\extracolsep{\fill}}lrrrrr}
\toprule 
 & \multicolumn{1}{c}{} &  & \multicolumn{3}{c}{SEE ($p$-value)}\tabularnewline
\cmidrule{4-6} \cmidrule{5-6} \cmidrule{6-6} 
 & Estimate &  & Standard & Bootstrap & Cluster bootstrap\tabularnewline
\midrule
$\theta_{0}$ & $.0089$ &  & $.0072$ $\left(.216\right)$ & $.0072$ $\left(.218\right)$ & $.0060$ $\left(.137\right)$\tabularnewline
$\theta_{1}$ & $.0533$ &  & $.0093$ $\left(.000\right)$ & $.0092$ $\left(.000\right)$ & $.0071$ $\left(.000\right)$\tabularnewline
$\theta_{2}$ & $-.0003$ &  & $.0160$ $\left(.984\right)$ & $.0164$ $\left(.984\right)$ & $.0147$ $\left(.982\right)$\tabularnewline
$\theta_{3}$ & $-.1227$ &  & $.0204$ $\left(.000\right)$ & $.0208$ $\left(.000\right)$ & $.0189$ $\left(.000\right)$\tabularnewline
\midrule
Wald $\chi^{2}\left(3\right)$ &  &  & $105.68$ & $112.09$ & $135.15$\tabularnewline
Model $p$ &  &  & $.0000$ & $.0000$ & $.0000$\tabularnewline
\bottomrule
\end{tabular*}
\par\end{centering}
\centering{}\textit{\footnotesize{}\medskip{}
}\\
\textit{\scriptsize{}Table notes:}{\scriptsize{} Each regression has
1,510 observations, log pseudo-$\mathcal{L}$ of $-1045.3992$, and
a pseudo-$R^{2}$ of $.0011$. Bootstraps involve 1,000 replications
each. There are $4\times6\times5=120$ feasible clusters of which
86 are populated. For the standard errors of estimates, the $p$-value
of a Wald test of the null hypothesis that the coefficient estimate
is equal to zero is reported in parantheses. The bottom row reports
the $p$-value of a Wald test against the null hypothesis that all
coefficient estimates are jointly zero. Rejections are ``highly significant''
for $p<.01$, ``significant'' for $p<.05$, and ``marginally significant''
for $p<.1$.}{\scriptsize\par}
\end{table}
\begin{table}[p]
\caption{Maximum likelihood estimates and standard errors of estimates (SEE)
of the non-linear belief updating model \eqref{eq:BeliefGLM} in beta
probit regression specification with transformation $2^{-1}\cdot\left(\Delta b+1\right)$.\label{tab:BeliefsBeta}}

\begin{centering}
\begin{tabular*}{1\textwidth}{@{\extracolsep{\fill}}lrrrrr}
\toprule 
 & \multicolumn{1}{c}{} &  & \multicolumn{3}{c}{SEE ($p$-value)}\tabularnewline
\cmidrule{4-6} \cmidrule{5-6} \cmidrule{6-6} 
 & Estimate &  & Standard & Bootstrap & Cluster bootstrap\tabularnewline
\midrule
$\theta_{0}$ & $.0092$ &  & $.0103$ $\left(.368\right)$ & $.0073$ $\left(.206\right)$ & $.0061$ $\left(.129\right)$\tabularnewline
$\theta_{1}$ & $.0532$ &  & $.0125$ $\left(.000\right)$ & $.0094$ $\left(.000\right)$ & $.0073$ $\left(.000\right)$\tabularnewline
$\theta_{2}$ & $.0010$ &  & $.0168$ $\left(.954\right)$ & $.0167$ $\left(.954\right)$ & $.0150$ $\left(.949\right)$\tabularnewline
$\theta_{3}$ & $-.1249$ &  & $.0203$ $\left(.000\right)$ & $.0217$ $\left(.000\right)$ & $.0198$ $\left(.000\right)$\tabularnewline
\midrule
Scale & $3.874$ &  & $.0360$ $\left(.000\right)$ & $.0814$ $\left(.000\right)$ & $.0794$ $\left(.000\right)$\tabularnewline
\midrule
Wald $\chi^{2}\left(3\right)$ &  &  & $113.54$ & $101.68$ & $116.39$\tabularnewline
Model $p$ &  &  & $.0000$ & $.0000$ & $.0000$\tabularnewline
\bottomrule
\end{tabular*}
\par\end{centering}
\centering{}\textit{\footnotesize{}\medskip{}
}\\
\textit{\scriptsize{}Table notes:}{\scriptsize{} The link function
is probit and the slink function log. Each regression has 1,510 observations
and log $\mathcal{L}$ of $1845.9897$. Bootstraps involve 1,000 replications
each. There are $4\times6\times5=120$ feasible clusters of which
86 are populated. For the standard errors of estimates, the $p$-value
of a Wald test of the null hypothesis that the coefficient estimate
is equal to zero is reported in parantheses. The bottom row reports
the $p$-value of a Wald test against the null hypothesis that all
coefficient estimates are jointly zero. Rejections are ``highly significant''
for $p<.01$, ``significant'' for $p<.05$, and ``marginally significant''
for $p<.1$.}{\scriptsize\par}
\end{table}

\begin{table}[p]
\caption{Maximum likelihood estimates and standard errors of estimates (SEE)
of the probit participation model \eqref{eq:ProbitModelAugm2} including
location and survey date fixed effects.\label{tab:ParticipationCF}}

\begin{centering}
\begin{tabular*}{1\textwidth}{@{\extracolsep{\fill}}lrrrr}
\toprule 
 & \multicolumn{1}{c}{} & \multicolumn{3}{c}{SEE ($p$-value)}\tabularnewline
\cmidrule{3-5} \cmidrule{4-5} \cmidrule{5-5} 
 & Estimate & Standard & Bootstrap & Cluster bootstrap\tabularnewline
\midrule
$\alpha$ & $-1.0855$ & $.1234$ $\left(.000\right)$ & $.1254$ $\left(.000\right)$ & $.1413$ $\left(.000\right)$\tabularnewline
$\beta$ & $-3.3062$ & $.7984$ $\left(.000\right)$ & $.8424$ $\left(.000\right)$ & $.7522$ $\left(.000\right)$\tabularnewline
\midrule 
Hamburg & $.1203$ & $.1084$ $\left(.267\right)$ & $.1104$ $\left(.276\right)$ & $.1231$ $\left(.328\right)$\tabularnewline
Munich & $.0827$ & $.1224$ $\left(.499\right)$ & $.1271$ $\left(.515\right)$ & $.1621$ $\left(.610\right)$\tabularnewline
Cologne & $.2485$ & $.1108$ $\left(.025\right)$ & $.1146$ $\left(.030\right)$ & $.1092$ $\left(.023\right)$\tabularnewline
\midrule 
S1 on Sep 9 & $-.0110$ & $.0944$ $\left(.907\right)$ & $.0947$ $\left(.907\right)$ & $.1055$ $\left(.917\right)$\tabularnewline
S1 on Sep 11 & $.0650$ & $.1357$ $\left(.632\right)$ & $.1471$ $\left(.659\right)$ & $.1671$ $\left(.697\right)$\tabularnewline
S1 on Sep 7 & $-.2655$ & $.1728$ $\left(.124\right)$ & $.1832$ $\left(.147\right)$ & $.1465$ $\left(.070\right)$\tabularnewline
S1 on Sep 6 & $-.0052$ & $.2259$ $\left(.982\right)$ & $.2365$ $\left(.983\right)$ & $.2692$ $\left(.985\right)$\tabularnewline
S1 on Sep 8 & $-.2767$ & $.3310$ $\left(.403\right)$ & $.2649$ $\left(.296\right)$ & $.2418$ $\left(.253\right)$\tabularnewline
\midrule 
S2 on Sep 17 & $-.0478$ & $.0957$ $\left(.618\right)$ & $.0913$ $\left(.601\right)$ & $.1156$ $\left(.679\right)$\tabularnewline
S2 on Sep 16 & $-.2168$ & $.1209$ $\left(.073\right)$ & $.1205$ $\left(.072\right)$ & $.1228$ $\left(.077\right)$\tabularnewline
S2 on Sep 19/20 & $-.1817$ & $.1720$ $\left(.291\right)$ & $.1817$ $\left(.318\right)$ & $.1324$ $\left(.170\right)$\tabularnewline
\midrule 
Corr. of errors & $.4519$ & $.1147$ & $.1237$ & $.1091$\tabularnewline
$\hat{e}$ st. dev. & $.1375$ & $.0025$ & $.0049$ & $.0045$\tabularnewline
\midrule 
Joint Wald $\chi^{2}\left(12\right)$ &  & $38.39$ & $39.95$ & $66.33$\tabularnewline
Joint $p$ &  & $.0001$ & $.0001$ & $.0000$\tabularnewline
\bottomrule
\end{tabular*}
\par\end{centering}
\centering{}\textit{\footnotesize{}\medskip{}
}\\
\textit{\scriptsize{}Table notes:}{\scriptsize{} Each regression has
1,510 observations and log $\mathcal{L}=344.82716$. $\Delta b$ is
instrumented by $z$, $c$, and $z\cdot c$. Berlin is the reference
category for location. The largest cells (Sep 10 for the first survey,
Sep 18 for the second survey) are reference categories for the survey
date indicators. The remaining survey date indicators are sorted by
cell size with the second-largest cell at the top, respectively. Sep
19 and Sep 20 are merged because the small number of observations
at Sep 20 predict success perfectly. A Wald test against the null
hypothesis that $\Delta b$ is exogenous (i. e. corr. of errors is
zero) is rejected with $\chi^{2}\left(1\right)=11.62$ and $p=.0007$.
Bootstraps call 1,000 replications each, with only complete ones being
used to calculate standard errors (135 replications failed in the
pooled specification and 153 in the clustered specification). There
are $4\times6\times5=120$ feasible clusters of which 86 are populated.
For the standard errors of estimates, the $p$-value of a Wald test
of the null hypothesis that the coefficient estimate is equal to zero
is reported in parentheses. The bottom row reports the $p$-value
of a Wald test against the null hypothesis that all coefficient estimates
are jointly zero. Rejections are ``highly significant'' for $p<.01$,
``significant'' for $p<.05$, and ``marginally significant'' for
$p<.1$.}{\scriptsize\par}
\end{table}
\begin{table}[p]
\caption{Newey's efficient minimum $\chi^{2}$ estimates and standard errors
of estimates (SEE) of the probit participation model \eqref{eq:ProbitModelAugm3}
including location and survey date fixed effects.\label{tab:ParticipationTS}}

\begin{centering}
\begin{tabular*}{1\textwidth}{@{\extracolsep{\fill}}lrrrr}
\toprule 
 & \multicolumn{1}{c}{} & \multicolumn{3}{c}{SEE ($p$-value)}\tabularnewline
\cmidrule{3-5} \cmidrule{4-5} \cmidrule{5-5} 
 & Estimate & Standard & Bootstrap & Cluster bootstrap\tabularnewline
\midrule
$\alpha$ & $-1.2163$ & $.1165$ $\left(.000\right)$ & $.1135$ $\left(.000\right)$ & $.1409$ $\left(.000\right)$\tabularnewline
$\beta$ & $-3.6764$ & $1.1187$ $\left(.001\right)$ & $1.1768$ $\left(.002\right)$ & $1.0265$ $\left(.000\right)$\tabularnewline
\midrule 
Hamburg & $.1347$ & $.1219$ $\left(.269\right)$ & $.1245$ $\left(.279\right)$ & $.1369$ $\left(.325\right)$\tabularnewline
Munich & $.0923$ & $.1379$ $\left(.503\right)$ & $.1444$ $\left(.523\right)$ & $.1832$ $\left(.614\right)$\tabularnewline
Cologne & $.2782$ & $.1237$ $\left(.024\right)$ & $.1276$ $\left(.029\right)$ & $.1193$ $\left(.020\right)$\tabularnewline
\midrule 
S1 on Sep 9 & $-.0124$ & $.1058$ $\left(.907\right)$ & $.1064$ $\left(.907\right)$ & $.1177$ $\left(.916\right)$\tabularnewline
S1 on Sep 11 & $.0731$ & $.1517$ $\left(.630\right)$ & $.1652$ $\left(.658\right)$ & $.1867$ $\left(.695\right)$\tabularnewline
S1 on Sep 7 & $-.2973$ & $.1925$ $\left(.123\right)$ & $.2056$ $\left(.148\right)$ & $.1629$ $\left(.068\right)$\tabularnewline
S1 on Sep 6 & $-.0058$ & $.2534$ $\left(.982\right)$ & $.2680$ $\left(.983\right)$ & $.3000$ $\left(.984\right)$\tabularnewline
S1 on Sep 8 & $-.3100$ & $.3735$ $\left(.407\right)$ & $.2959$ $\left(.295\right)$ & $.2696$ $\left(.250\right)$\tabularnewline
\midrule 
S2 on Sep 17 & $-.0535$ & $.1072$ $\left(.618\right)$ & $.1022$ $\left(.601\right)$ & $.1280$ $\left(.676\right)$\tabularnewline
S2 on Sep 16 & $-.2426$ & $.1353$ $\left(.073\right)$ & $.1354$ $\left(.073\right)$ & $.1381$ $\left(.079\right)$\tabularnewline
S2 on Sep 19/20 & $-.2034$ & $.1921$ $\left(.290\right)$ & $.2037$ $\left(.318\right)$ & $.1453$ $\left(.162\right)$\tabularnewline
\midrule 
Joint Wald $\chi^{2}\left(12\right)$ &  & $24.33$ & $25.33$ & $44.27$\tabularnewline
Joint $p$ &  & $.0183$ & $.0133$ & $.0000$\tabularnewline
\bottomrule
\end{tabular*}
\par\end{centering}
\centering{}\textit{\footnotesize{}\medskip{}
}\\
\textit{\scriptsize{}Table notes:}{\scriptsize{} Each regression has
1,510 observations. $\Delta b$ is instrumented by $z$, $c$, and
$z\cdot c$. Berlin is the reference category for location. The largest
cells (Sep 10 for the first survey, Sep 18 for the second survey)
are reference categories for the survey date indicators. The remaining
survey date indicators are sorted by cell size with the second-largest
cell at the top, respectively. Sep 19 and Sep 20 are merged because
the small number of observations at Sep 20 predict success perfectly.
A Wald test against the null hypothesis that $\Delta b$ is exogenous
(i. e. corr. of errors is zero) is rejected with $\chi^{2}\left(1\right)=10.82$
and $p=.0010$. Bootstraps call 1,000 replications each, with only
complete ones being used to calculate standard errors (135 replications
failed in the pooled specification and 153 in the clustered specification).
There are $4\times6\times5=120$ feasible clusters of which 86 are
populated. For the standard errors of estimates, the $p$-value of
a Wald test of the null hypothesis that the coefficient estimate is
equal to zero is reported in parentheses. The bottom row reports the
$p$-value of a Wald test against the null hypothesis that all coefficient
estimates are jointly zero. Rejections are ``highly significant''
for $p<.01$, ``significant'' for $p<.05$, and ``marginally significant''
for $p<.1$.}{\scriptsize\par}
\end{table}

\end{document}